\documentclass[sigconf,screen]{acmart}

\usepackage{lscape}
\usepackage{float}
\usepackage{xcolor}
\usepackage{multirow}
\usepackage{graphicx}
\usepackage[toc,page]{appendix}
\usepackage{stfloats}
\usepackage{pdflscape}

\AtBeginDocument{%
  \providecommand\BibTeX{{%
    \normalfont B\kern-0.5em{\scshape i\kern-0.25em b}\kern-0.8em\TeX}}}

\copyrightyear{2025}
\acmYear{2025}
\copyrightyear{2025}
\acmConference[FAccT '25]{The 2025 ACM Conference on Fairness, Accountability, and Transparency}{June 23--26, 2025}{Athens, Greece}
\acmDOI{10.1145/3715275.3732069}
\acmISBN{979-8-4007-1482-5/2025/06}

\begin{document}

\title[Stakeholder Participation for Responsible AI Development]{Stakeholder Participation for Responsible AI Development: Disconnects Between Guidance and Current Practice}

\author{Emma Kallina}
\orcid{0000-0003-4912-7216}
\affiliation{%
  \department{Research Centre Trust}
  \institution{University-Alliance Ruhr}
    \city{University Duisburg-Essen}
  \country{Germany}
}
\affiliation{%
  \institution{University of Cambridge}
  \city{Cambridge}
  \country{United Kingdom}
}

\author{Thomas Bohn\'e}
\orcid{0000-0001-5986-8638}
\affiliation{%
  \institution{University of Cambridge}
  \city{Cambridge}
  \country{UK}
}

\author{Jatinder Singh}
\orcid{0000-0002-5102-6564}
\affiliation{%
  \department{Research Centre Trust}
  \institution{University-Alliance Ruhr}
    \city{University Duisburg-Essen}
  \country{Germany}
}
\affiliation{%
  \institution{University of Cambridge}
  \city{Cambridge}
  \country{United Kingdom}
}

\begin{abstract}

Responsible AI (rAI) guidance increasingly promotes stakeholder involvement (SHI) during AI development.
At the same time, SHI is already common in commercial software development, but with potentially different foci. 
This study clarifies the extent to which established SHI practices are able to contribute to rAI efforts as well as potential disconnects -- essential insights to inform and tailor future interventions that further shift industry practice towards rAI efforts.
First, we analysed 56 rAI guidance documents to identify why SHI is recommended (i.e. its expected benefits for rAI) and uncovered goals such as redistributing power, improving socio-technical understandings, anticipating risks, and enhancing public oversight.
To understand why and how SHI is currently practised in commercial settings, we then conducted an online survey (n=130) and semi-structured interviews (n=10) with AI practitioners. 
Our findings reveal that SHI in practice is primarily driven by commercial priorities (e.g. customer value, compliance) and several factors currently discourage more rAI-aligned SHI practices.
This suggests that established SHI practices are largely not contributing to rAI efforts. 
To address this disconnect, we propose interventions and research opportunities to advance rAI development in practice.

\end{abstract}



\maketitle

\section{Introduction} \label{sec:intro}
Policymakers, regulators, civil society, and researchers are dedicating significant efforts to advance more responsible AI (rAI) practices. \textbf{A growing body of work emphasises stakeholder involvement (SHI) throughout the AI lifecycle as an important component of these rAI efforts} \citep[e.g.][]{stray2020aligning, Turing_Doc, lu2022responsible, aizenberg2020designing, sadek2024challenges, kallina2024stakeholder}. 
In the software development context \textit{stakeholders} are described as an ``[i]ndividual or organization [...] having a right, share, claim, or interest in a system or in its possession of characteristics that meet their needs and expectations'' \citep[\S3.44, ISO 15288,][]{ISO2}. These stakeholders can be internal or external to the organisation developing the AI system, and identifying the relevant stakeholders of an AI system can be complex \citep{miller2022stakeholder, lu2022responsible}. Internal stakeholders can include engineers, the legal team, and management, while external stakeholders encompass users, affected individuals, at-risk groups, and third-party auditors to name but a few \cite{miller2022stakeholder}.
SHI describes engaging these stakeholders at various points in the AI lifecycle to understand their needs, values, and concerns in the application context to then address these in the system’s design \citep{shah2022stakeholder}.

Reflecting the recent emphasis on SHI in the rAI literature \cite[e.g.][]{kallina2024stakeholder, sadek2024challenges, birhane2022power, corbett2023power, shah2022stakeholder, lu2022responsible}, many global organisations have incorporated SHI into their rAI guidance documents, designed to encourage organisations to adopt and navigate rAI development practices -- and thus potentially influencing future AI policies and regulations (\S\ref{sec:AIpolicy}). These organisations include multi-stakeholder and intergovernmental organisations (e.g. ACM \cite{ACM_Doc_2, ACM_Doc_1}, Partnership on AI \cite{PartnerSHIpAI_Doc1, PartnerSHIpAI_Doc2}; OECD \cite{OECD_Doc}, UN \cite{UN_Doc}), government bodies (e.g. White House Office for Science and Technology \cite{US_Doc, US_Doc2, US_Doc3}, European Commission \cite{HLEGAI_EC}), and civil society organisations (e.g. Algorithm Watch \cite{AlgorithmWatch_Doc1, AlgorithmWatch_Doc2}, Amnesty International \cite{Amnesty_2, Amnesty_Doc1, Amnesty_Doc3}). 

Independent from these calls, \textbf{SHI is already a well-established practice in commercial software development}, playing a key role in methodologies such as agile or user-centred design (\S\ref{sec:traditional}). This raises the question of whether or to what extent these `traditional', currently common forms of SHI in industry (\S\ref{sec:traditional}) are already contributing to the benefits associated with SHI in rAI guidance documents. Clarifying this is essential to inform effective measures and interventions that promote SHI practices which align with and support rAI efforts.

\noindent To clarify \textbf{whether current SHI practices in industry already contribute to rAI efforts}, we explore the following: 

\begin{description}
    \item \textbf{(Q1)} which positive outcomes does rAI guidance associate with\slash desire from recommending SHI during AI development?
    \item \textbf{(Q2)} what are the currently established SHI practices in commercial AI development settings?
    \item \textbf{(Q3)} whether and to what extent are these currently established SHI practices in line with and able to contribute to the positive outcomes that rAI guidance associates with SHI?
\end{description}

To address these questions, we conducted a mixed-method study.
For Q1, we reviewed and analysed 56 AI guidance documents from 29 organisations (\S\ref{sec:AIpolicy}) to identify the benefits that these organisations associate with SHI in AI development (\S\ref{sec:benefits}).
Exploring Q2, we conducted an online survey with 130 AI practitioners and deep-dive semi-structured interviews (n = 10) to gather insights into currently established SHI practices in commercial settings. 
To identify whether\slash the extent to which these contribute to the five benefits identified in Q1, we compared our findings from Q1 and Q2, thereby addressing Q3. Our findings revealed a strong disconnect: \textbf{current SHI practices reflect traditional SHI practices which are largely not in line with rAI efforts}, and thus fail to achieve more responsible practices (\S\ref{sec:comparing}). Further, we uncovered conflicts between commercial interests and rAI-aligned SHI, thus discouraging SHI beyond this `traditional' scope (\S\ref{sec:SHI_conflictcommercial}). 
Based on these findings, we then derive interventions and opportunities for further research that address this disconnect and advance SHI for rAI efforts in the commercial practice (\S\ref{sec:recommendations}).
\noindent In short, this study offers following \textbf{four key contributions} to the rAI community:
\small
\begin{itemize}
\item \textbf{Deriving and clarifying the benefits associated with SHI for rAI development}. This knowledge provides a foundation for the rAI community to reflect on, create, or refine interventions or tools that support the achievement of these benefits as well as to assess their fulfilment or whether there are further aspects they wish to cover. This clarity further supports industry practitioners in reflections about their own drivers for their SHI efforts, identifying potential misalignments. 
\item \textbf{Identifying the alignment of established SHI practices with rAI efforts.} These findings enable the rAI research community to shape an immediate research agenda towards investigating and realising rAI-aligned SHI practices (§\ref{sec:researchagenda}). Further, the findings facilitate the creation and adoption of more effective rAI guidance by industry practitioners and organisations in the AI and rAI space.
\item \textbf{Detailing the tensions discouraging SHI aligned with rAI guidance}, i.e. currently contradictory incentive structures between the commercial practice and rAI efforts. These insights are essential to scope further research and interventions from both the rAI research community and industry to shift current practices towards rAI goals. Moreover, this information enables AI policy makers and regulators to tailor their regulations and guiding actions towards addressing these tensions, i.e. to better incentivise SHI that is in line with rAI efforts. This becomes particularly relevant considering the emergence of AI regulation that encourages organisations to engage in SHI to advance rAI goals \citep[see e.g.][]{US_Doc, EUAIAct}.
\item \textbf{Indicating actions to advance rAI-aligned SHI in the AI practice}. By outlining actionable recommendations (§\ref{sec:recommendations}) and a clear research agenda (\S\ref{sec:researchagenda}), we offer concrete starting points to overcome current challenges and to further align SHI practices with rAI goals.
\end{itemize}
\normalsize

\section{Background} \label{sec:background}
The following provides an overview over key issues with rAI guidance in general, the traditional forms of SHI in software development, and insights into currently established SHI practices.

\subsection{Issues With rAI Guidance in General} 
\label{sec:otherguidance}

rAI guidance aims to steer the development of AI systems to prevent negative outcomes for the public and especially marginalised communities. The majority of such guidance aims at AI practitioners, however, is often developed with limited input from the commercial practice (thus potentially failing to account their perspective and practice), or published behind paywalls \citep{holstein2019improving, chivukula2021surveying}. 

Recent studies have engaged with AI practitioners to understand their perception and use of such guidance, e.g. for AI fairness \cite{madaio2020co}, rAI principles \cite{rakova2021responsible, mcnamara2018ACMcode, vakkuri2019ethically}, and design standards for more transparent systems \cite{schor2024standards}. Across these studies, \textbf{practitioners reported difficulties when applying such rAI guidance\slash tools to their own projects or workplace}, suggesting that similar obstacles might emerge when applying SHI-related guidance.
Many of the reported challenges reflect broader critiques of rAI guidance, voiced by academics and practitioners.
A common concern is that rAI guidance is too abstract to be actionable in practice \cite{rakova2021responsible, mittelstadt2019principles, hagendorff2020ethics, schor2024standards, sadek2024challenges} and it has been shown that awareness of rAI principles alone fails to influence practitioner actions \cite{mcnamara2018ACMcode}. Other criticisms include the inherent subjectivity of rAI principles such as fairness or transparency (i.e. where objective measures are difficult) \cite{sadek2024challenges, schor2024standards} and that they represent predominantly Western views \cite{jobin2019global}.

Our paper differs from such work by focusing not on critiquing rAI guidance documents, but on clarifying the extent to which established SHI practices in industry align with the SHI-related goals of such rAI guidance. This knowledge is essential to improve this alignment going forward, e.g. through more actionable support.

\subsection{SHI’s Traditional Role in Software Creation}\label{sec:traditional}
SHI has been an integral part of software development for decades, from early value-based software engineering \cite[e.g.][]{biffl2006value, roberts2000perceptions} to the two most dominant methods during software development and design today: agile and user-centred design which can be used individually or combined  \cite[][]{jurca2014integrating, salah2014systematic}.

\textit{Agile} is the currently most common software development technique used in industry \citep{StateofAgile, jurca2014integrating}. The influential Agile Manifesto \citep{Agile_Manifesto} (2001) values ``customer collaboration over contract negotiation'', promoting the iterative discovery of system requirements in collaboration with the customer (vs. static, \textit{a priori} defined system requirements). Thus, the close collaboration with end-users and customers is promoted in order to maximise customer value and thus the financial success of a product \cite{Agile_Wiki}.
\textit{User-centred design} focuses on ``provid[ing] the best user experience possible'' \citep[p. 5399,][]{chammas2015closer} through the continuous involvement of potential users during the system's design \citep{chammas2015closer, karat1996user, gulliksen2003key}. This popular approach became formalised in the ISO standard 9241-210 (2010) \citep{ISO} and focuses primarily on the interface.

Both prominent approaches entail SHI, suggesting that SHI is already an established practice in software development. However, the narrow focus on customer value or the user interface might imply that \textbf{only stakeholders that are tied to commercial interests or direct interactions with the interface are involved}. This limited scope may still prevail in established SHI practices.

\subsection{Current SHI Practices in Academia \& Commercial Settings} \label{sec:currentinsights}
Considering the history of SHI in software development, we require insights into how these practices are deployed during AI development today. Two recent studies analyse external SHI in \textit{academic} AI development \citep{delgado2023participatory, corbett2023power}. The first study analysed a corpus of 80 research articles reporting the development of software with external SHI \citep{delgado2023participatory}, demonstrating that the \textbf{majority of studies involved stakeholders without empowering them to influence core objectives} (i.e. low decision-authority). \citet{corbett2023power} found a similar pattern after mapping recent academic work reporting the development of an AI system to Arnstein's Ladder of Citizen Participation \citep[][]{arnstein1969ladder}. \citet{greenwood2007stakeholder} emphasises that SHI without granting high levels of decision-authority must not be automatically considered as advancing ethical goals -- and thus rAI.

Due to its novelty and emerging nature, the \textit{commercial} practice of SHI during AI development is \textbf{not well-studied}. An exception is \citet{groves2023going} who investigated AI practitioners' awareness of and attitudes towards public participation, a type of SHI that aims at involving members of the public beyond domain experts or users. They found that practitioners had theoretical knowledge of best practices but rarely reported instances of public participation at their workplace. The participants---selected based on their positive attitude towards SHI---reported barriers to establishing such practices, often related to associated costs (time \& resources). The authors conclude that public participation in industry is rare and that the general buy-in to invest resources is low, even when individual employees are promoting this approach (prevalence and knowledge in a less SHI-attuned sample might be far lower). 
These first insights hint towards the fact that \textbf{established SHI efforts rarely include members of the public}. To advance this understanding, we require insights into SHI practices beyond public participation, focusing on concrete experiences (vs theoretical knowledge). This enables us to derive the extent to which established SHI practices are in line with rAI efforts -- the goal of this study.

\section{Method} \label{sec:methods}
We now describe our three-step multi-method approach, designed to address the questions outlined in \S\ref{sec:intro}. 
First, we analysed the benefits that rAI guidance associates with SHI by reviewing 56 documents from organisations issuing AI guidance (Q1).
To investigate the status quo of SHI in commercial AI development (Q2), we conducted an online survey with AI practitioners which uncovered a broad range of current practices, drivers, and attitudes, before conducting semi-structured interviews to dive deeper into identified patterns. Then, we related the findings from Q1 and Q2 to derive the extent to which currently established SHI practices are able to achieve these benefits (Q3). 

\subsection{Analysis of Official AI Guidance: Drivers Behind Encouraging SHI}\label{sec:AIpolicy}
To understand the benefits that rAI guidance associates with SHI during AI development, we reviewed a broad range of rAI guidance documents. These documents seek to shape industry practice by providing recommendations around the development and deployment of AI systems, usually with the goal of preventing harmful outcomes. Considering the numerous recent scandals caused by AI systems \citep[e.g.][]{AI_Incident_Database}, industry practitioners are likely to seek such guidance to minimise risks as well as to manage their public image.
Additionally, while these documents are typically not themselves binding, upcoming AI regulations are already indirectly pointing towards these documents: by often stating only general, high-level requirements (e.g. to follow ``an inclusive and diverse design [process] of AI systems, including through the [...] promotion of stakeholders’ participation", Chapter 10, EU AI Act \cite{EUAIAct}) they leave the need for more concrete and actionable guidance. Thus, AI practitioners might consult rAI guidance for more detailed recommendations as well as to demonstrate compliance when audited.
Furthermore, these documents are a probable starting point for best practices within organisations and may influence the development of future regulations -- or at least inform the iterative process of establishing converging and actionable best practices that regulatory measures seek to encourage. 
Taken together, guidance documents such as the ones analysed in \S\ref{sec:benefits} are \textbf{highly likely to have direct or indirect legal and compliance implications}. Thus, it is essential to understand discrepancies between such guidance and the currently established commercial practice: the aim of this study.

\subsubsection{Materials \& Analysis}
We compiled a list of organisations publishing rAI guidance given we did not discover a peer-reviewed list of such. Our key aim was to include a broad range of organisations to identify common reasons for recommending SHI for rAI development. %
The resulting list included 29 organisations (\S\ref{sec:benefits}). We reviewed all available rAI guidance publications on each organisations' websites and retrieved 56 guidance documents or recommendations regarding SHI during the development or governance of AI systems.
For this final list, we excluded rAI principles or guidance from private sector organisations. Our aim was to compare the goals of non-commercial rAI actors with the commercial, private sector practice. Thus, including documents issued by private sector organisations might obscure this comparison, e.g. due to conflicts of interests.
We systematically analysed the relevant 56 publications, using both skimming technique \cite{dhillon2020effect} as well as a set of keywords associated with SHI (i.e. likely to identify relevant paragraphs). On relevant paragraphs, we then performed a manual thematic analysis following \citet{braun2012thematic} using the software Atlas.ti \citep{Atlasti}.
See Appx \ref{appendix:orglist} for further details on this process.

\subsection{Practitioner Insights: Survey \& Interviews}\label{sec:practitioner}
To gain insights into \textit{current} SHI practices in commercial AI settings, we combined an anonymous online survey with semi-structured interviews with AI practitioners. This work was approved by our department's ethical review board.

\subsubsection{Materials \& Procedure}
The online survey allowed us to gather a broad picture of current SHI practices in industry. Participants were asked to answer most questions in relation to one specific system they currently work on\slash have worked on in the past, ensuring that they reported concrete experiences rather than best practices or attitudes \cite[as in][]{groves2023going}. 
The semi-structured interviews aimed at uncovering the mechanisms underlying the identified practices and patterns as well as to gather rich qualitative insights and illustrative quotes. By reviewing the participant's survey answers beforehand, we could delve deeper into underlying reasons and cover broader tensions\slash high-level considerations, both on general matters as well as specific responses.
The remote interview sessions were conducted with Microsoft Teams \citep{MS_Teams}, lasted between 56 and 81 minutes, and were recorded with automatic transcription. The semi-structured interview guide and the survey questions can be viewed our supplementary materials.

\subsubsection{Recruitment \& Sample}
We recruited survey participants via convenience sampling the researchers' online networks (LinkedIn, Twitter, Reddit) \citep{sedgwick2013convenience} and the platform Prolific \citep{Prolific} through a pool of research participants. To ensure that our participants had experience along the AI pipeline we took various actions: firstly, we used available filter criteria related to technology design to reach participants with relevant backgrounds (e.g. `coding' OR `UI design' OR `A/B Testing'; filtering for AI practitioners directly was not possible). Further, we explicitly stated the prerequisite of being an AI practitioner in the survey description and included filter questions about AI experience that aborted the survey if not met. Additionally, we conducted pilot trials with 20 participants to assess submission quality. Given the exploratory nature of the study, we cast a deliberately wide net and explored SHI practices across roles and industries.
Participants were paid based on the local living wage and the median completion time was 22 minutes.
Of the 149 participants, 17 did not finish the survey and were excluded, together with two further participants with an unrealistically short completion time (under four minutes). 99 of the remaining 130 participants were recruited via Prolific. Appx. \ref{appendix:sample} provides more details on our participants.
Since the interviews sought to dive deeper into identified practices, we recruited participants through their survey responses, i.e. invited participants who voluntarily provided their email on the last page of the survey to be involved in further studies on the topic. This led to ten interviews, reflecting the difficulties of recruiting an expert sample for extended periods of time (\S\ref{sec:limitations}).

\subsubsection{Analysis} 
The numeric survey data (e.g. frequency with which a specific answer was selected) was analysed using Python \citep{Python}, detailed statistics in Appx. \ref{appendix:stats}.
We analysed our qualitative data following a thematic analysis approach \cite{braun2012thematic}, i.e. the iterative coding of the interview transcripts and the open text entries of the survey to derive subthemes, themes, and ultimately patterns of meaning. 

\section{Results}

\subsection{RAI Guidance: Five Key Benefits of SHI}

\label{sec:benefits}
Addressing Q1, we reviewed rAI guidance documents to identify the benefits they associate with SHI.
The thematic analysis of 56 guidance documents issued by 29 organisations---eleven multi-stakeholder organisations (e.g. ACM, Partnership on AI), eight intergovernmental organisations (e.g. OECD, UN, WEF), four civil society organisations (e.g. Algorithm Watch, Centre for Democracy \& Technology), and a selection of six governmental organisations---resulted in 15 codes related to the benefits of SHI. 
These were grouped into five themes, i.e. the key benefits that rAI guidance associates with SHI: (1) the rebalancing of decision power, (2) an improved understandings of the system's socio-technical context, (3) a better anticipation of risks, (4) increased understanding and trust, and (5) to enable public scrutiny and monitoring. The results are detailed in Table \ref{tab:codes} which presents the themes with their sub-codes, descriptions, and illustrative quotes. Taking these together, rAI guidance seem to \textbf{promote SHI as an instrument to shift the focus and agency during the AI development process towards affected communities}, beyond that of the organisations developing or commissioning AI systems. %

\subsection{Practitioner Insights: Current Practices Reflect Traditional Forms of SHI} \label{sec:practitioner_practices}
The survey and semi-structured interviews with AI practitioners allowed us to gain a broad overview over current drivers and practices of SHI, addressing Q2. We present our key findings from the survey and interviews jointly to offer more quantitative evidence on broader trends together with qualitative data and quotes. 
Throughout this section, quotes are marked as follows: \textit{`survey answer option'}, \textit{``survey participant quote''}, and ``interviewee quote''. Further graphs and statistical details can be viewed in Appx. \ref{appendix:surveyresults}.

\subsubsection{Commercial Interests are the Key Driver Behind SHI} 
\label{sec:SHI_drivers}
\begin{figure*}[t]
  \centering
  \includegraphics[width=.8\textwidth]{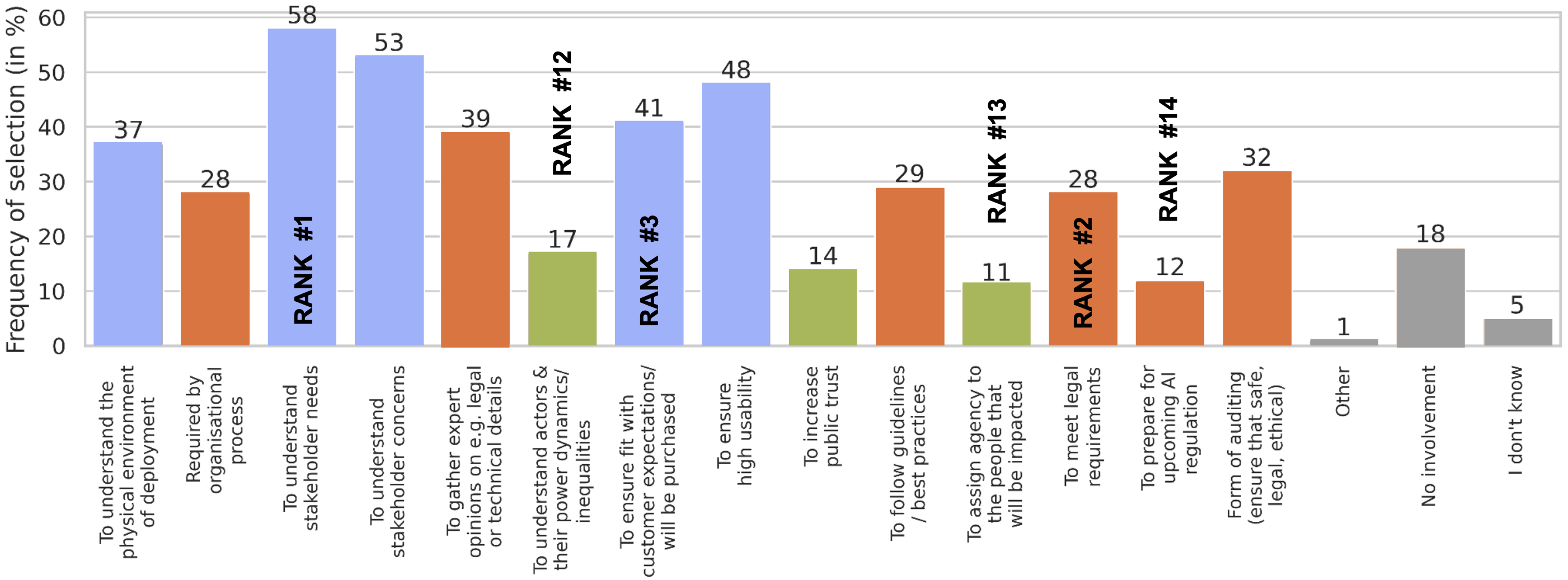}
  \caption{Key drivers of SHI as reported by survey participants. Colours illustrate the three broad categories discussed in \S\ref{sec:SHI_drivers}: \textcolor{blue}{\textbf{blue}} = traditional SHI drivers, \textcolor{orange}{\textbf{orange}} = SHI drivers related to compliance, \textcolor{olive}{\textbf{green}} = drivers beyond commercial \& compliance interests. The ranks indicate the top three highest and lowest ranks that participants assigned to their %
  selected drivers.}
  \label{fig:drivers}
  \Description{Bar chart illustrating how often the survey participants selected the various drivers. The colours of the bars is classifying the drivers of SHI in three categories. Drivers of the first category were selected with the highest frequency, drivers from the second category (i.e. compliance matters) were selected with a medium frequency. Drivers beyond traditional foci of SHI were selected very rarely.}
\end{figure*}

Our survey participants indicated that the most frequent motivations behind SHI during AI development resembled traditional drivers of SHI, i.e. to \textbf{increase customer value and\slash or usability} through identifying user or customer needs and expectations. As the blue bars in Fig. \ref{fig:drivers} show, these drivers were selected with the highest frequency, including \textit{`understanding stakeholders' needs'} or \textit{`concerns'} (selected by 58\% and 52\%), \textit{`increased usability'} (48\%) and a \textit{`better fit to customer expectations'} (41\%). 
This was confirmed by our interviewees: 90\% mentioned SHI as a method to ``ensur[e] that the client got something that was meeting their needs fully'' (I1) and ``to identify product-market fit'' (I10). Several interviewees further specified that the underlying goal was to ensure the system's financial viability, i.e. ``all they [i.e. my management] care about is the fact that they're going to buy it'' (I4).

Commercial interests further drove SHI for regulatory or internal compliance purposes (orange bars in Fig. \ref{fig:drivers}). Other than the traditional drivers that focused on product-specific commercial interests, these drivers describe \textbf{commercial interests related to legal or reputational risks}. While they were selected with a slightly lower frequency than traditional SHI drivers, they were highly important for practitioners that \textit{did} consider compliance aspects: the 28\% of survey participants that selected the driver \textit{`to meet legal requirements'} ranked it as second most important driver when ranking their previously selected drivers (ranking in Fig. \ref{fig:drivers}). Mirroring this, half of the interviewees reported legal considerations as motivation for SHI, e.g. the involvement of domain experts ``to give us their professional OK [...]. We didn't want to violate any legal limit'' (I5).

In contrast, \textbf{drivers reflecting goals beyond traditional SHI were selected more rarely}, i.e. drivers not directly linked to commercial interests (but instead to e.g. rAI efforts) such as \textit{`to assign agency to the people impacted by the system'} (11\%) or \textit{`to increase public trust'} (14\%). Even if selected, such drivers were ranked with the lowest importance (Fig. \ref{fig:drivers}, green bars). 
These findings suggests that commercial interests are the primary drivers behind SHI rather than those related to rAI efforts.

\begin{landscape}
\begin{table}[]
\caption{Themes resulting from the analysing the drivers behind the recommendation of SHI along the AI lifecycle in 56 AI guidance documents, together with the sub-codes and example quotes.}
\label{tab:codes}
\resizebox{\columnwidth}{!}{%
\begin{tabular}{llll}
\textbf{Theme\slash Benefit of SHI} &
  \textbf{Relevant Codes} &
  \textbf{Description} &
  \textbf{Example Quotes} \\ \hline
\begin{tabular}[c]{@{}l@{}}\textbf{Rebalance Decision} \\ \textbf{Power} \\ \\ (encouraged by \\ n = 26 organisations) \vspace{86pt}\end{tabular}&
  \begin{tabular}[c]{@{}l@{}}\textbf{inclusive / societal vision} of the type of AI should be \\ developed and which not, including who it should serve \\ \\ \textbf{redistribute agency} around how AI is \\ developed and designed \\ \\ \textbf{giving affected communities a voice} in the design \\ of systems impacting their lives \vspace{54pt} \end{tabular}&
  \begin{tabular}[c]{@{}l@{}}Refers to the capability of SHI to redistribute agency in decision processes \\ along the AI lifecycle\\ \\ Currently, the communities envisioning and creating future AI systems \\ reflects historical power imbalances \citep{hoffmann2019fairness, AINow}, for example, only 10\% of \\ AI research staff at Google is female and only 2.5\% is black \citep{AINow}). \\ Consequently, AI systems are created based on a narrow range of lived \\ experiences neglecting interests and visions that are not represented \citep{AINow}. \vspace{52pt} \end{tabular} &
  \begin{tabular}[c]{@{}l@{}}University of Montreal: ``citizens should have the opportunity and skills to deliberate on the \\ social parameters of these AIS [(i.e. AI systems)], their objectives, and the limits of their use'' \\ since these systems have ``a significant impact on the life of citizens'' \citep[p. 12,][]{Montreal}.\\ \\ Alan Turing Institute: SHI is required to ``take the reins of innovation and to steer the course \\ of our algorithmic creations in accordance with a shared vision of what a better human \\ future should look like'' \citep[p. 73,][]{Turing_Doc}{}.\\ \\ WHO for AI in the healthcare context that ``[p]atients, community organisations and civil \\ society should be able to hold governments and companies to account, to participate in the \\ design of technologies and rules, to develop new standards and approaches and to demand \\ and seek transparency to meet their own needs as well as those of their communities and \\ health systems'' \citep[p. 16,][]{WHO_Doc}.\end{tabular} \\ \hline
\begin{tabular}[c]{@{}l@{}}\textbf{Detailed} \\ \textbf{Understandings of} \\ \textbf{the Socio-Technical} \\ \textbf{Context} \\ \\ (encouraged by \\ n = 26 organisations) \vspace{55pt}\end{tabular}&
  \begin{tabular}[c]{@{}l@{}}\textbf{methods of achieving a deeper understanding} are \\ inviting both multiple professional experiences as well \\ as multiple lived experiences of affected communities, \\ emphasising that experiences and expertise beyond \\ technical expertise are required\\ \\ \textbf{increased fit with context} through involving actors \\ that will influence the expected application context \\ \\ \textbf{translating abstract principles to a use case}, \\ i.e. SHI as a step to understand how a facet of rAI \\ should manifest in a specific system \vspace{2pt} \end{tabular} &
  \begin{tabular}[c]{@{}l@{}}SHI leads to an improved understanding of a system's application context \\ and social aspects, enabling the creation of systems that fit to their \\ environment. Lacking such fit can lead to various downstream issues: \\ \citet{zajkac2023clinician}, e.g. found that the most issues when deployment ML \\ systems in the clinical practice was caused by an insufficient fit of the \\ systems with clinicians’ needs and existing clinical processes.\\ \\ The in-depth involvement allows affected communities to translate \\ abstract rAI principles (e.g. Fairness, Transparency, ...) to their specific \\ needs in a use case, harnessing their insights into existing \\ inequalities and imbalances. \vspace{10pt} \end{tabular} &
  \begin{tabular}[c]{@{}l@{}}Alan Turing Institute: ``[a]ll individual human beings come from unique places, experiences, \\ and life contexts that have shaped their thinking and perspectives. Reflecting on these is \\ important insofar as it can help team members understand how their viewpoints might differ \\ from those [...][with] diverging cultural and socioeconomic backgrounds and life experiences. \\ Identifying and probing these differences can enable individuals to better understand how \\ their own backgrounds [...] frame the way they see others, the way they approach and solve \\ problems [...]. The answers to these questions will always be tricky, but deliberation between \\ team members and stakeholders should always be a part of arriving to any sort of consensus'' \\ \citep{Turing_Course}.\\ \\ World Economic Forum: ``[rAI] principles should first be contextualised to reflect the local \\ values, social norms and behaviours of the community in which the AI solutions operate'' \citep[][]{WEF_Doc2}.\end{tabular} \\ \hline
\begin{tabular}[c]{@{}l@{}}\textbf{Improved Risk}\\ \textbf{Anticipation} \\ \\ (encouraged by \\ n = 22 organisations) \vspace{76pt}\end{tabular}&
  \begin{tabular}[c]{@{}l@{}}\textbf{more accurate formal impact assessment}\\ regarding harm identification pre-deployment\\ \\ \textbf{more expansive informal prediction of risks}\\ through more diverse perspectives\\ \\ \textbf{detect and mitigate discriminatory outcomes}\\ through including marginalised communities\\ \\ \textbf{understanding and resolving value tensions}\\ resulting from the diverse and often contradictory \\ goals of different stakeholders\end{tabular} &
  \begin{tabular}[c]{@{}l@{}}Related to the previous benefit of SHI, i.e. an improved understanding of a \\ system's context. This knowledge allows for a better anticipation of negative \\ consequences that a system might cause, as well as the evaluation of \\ proposed or implemented mitigation strategies. \\ \\ Furthermore, value tensions between conflicting goals of different \\ stakeholders (common in the context of AI systems, \citep[see e.g.][]{deshpande2022responsible} can be \\ addressed and mitigated through SHI as demonstrated in recent work \\ \citep[e.g.][]{kunkel2023more, lee2019webuildai} \vspace{30pt} \end{tabular} &
  \begin{tabular}[c]{@{}l@{}}White House Office of Science and Technology Policy: ``[a]utomated systems should be \\ developed with consultation from diverse communities, stakeholders, and domain experts \\ to identify concerns, risks, and potential impacts of the system'' \citep[p. 5,][]{US_Doc}, \\ engaging `diverse impacted communities to consider concerns and risks that may be unique \\ to those communities'' \citep[p. 18,][]{US_Doc}.\\ \\ Amnesty International: ``companies developing these technologies must take immediate \\ steps to proactively engage with academics, civil society actors, and community \\ organisations, especially those representing traditionally marginalised communities. \\ Although we cannot predict all the ways in which this new technology can and may cause \\ or contribute to harm, we have extensive evidence that marginalised communities are most \\ likely to suffer the consequences'' \citep{Amnesty_2}.\end{tabular} \\ \hline
\begin{tabular}[c]{@{}l@{}}\textbf{Increased Public}\\ \textbf{Understanding and} \\ \textbf{Trust}\\ \\ (encouraged by \\ n = 16 organisations)\\ \\ \vspace{26pt}\end{tabular} &
  \begin{tabular}[c]{@{}l@{}}\textbf{increased transparency as a pre-condition}, i.e. \\ it has to be transparent to stakeholders that and where \\ they are able to participate in the development process\\ \\ \textbf{increased transparency as a consequence} through \\ allowing various stakeholders to gain insights into the \\ development process and the system's objective\\ \\ \textbf{better understanding} of the capabilities of AI \\ might result in better calibrated trust \vspace{6pt} \end{tabular} &
  \begin{tabular}[c]{@{}l@{}}SHI can increase the transparency of the development process, demystifying \\ its opaque steps. This can help to increase the general understanding of \\ AI systems' capabilities and the development process itself. as well as the \\ capabilities, objectives, and the development of specific AI systems. \\ Especially the latter allows involved stakeholders to make informed \\ decisions regarding how they calibrate their trust towards the system. \vspace{45pt} \end{tabular} &
  \begin{tabular}[c]{@{}l@{}}Ada Lovelace Institute: SHI ``provides [...] greater legitimacy and accountability by ensuring \\ those who are affected have their voices and perspectives taken into account. [This] [...] can \\ support a [...] technology ecosystem that earns the trust of the public'' \citep[p. 73,][]{Ada_Doc2}.\\ \\ Data \& Society: ``Public participation adds legitimacy to decisions because people \\ trust processes they understand and influence'' \citep[p. 3,][]{DemocratisingAI}.\\ \\ World Economic Forum: ``demystif[ying] AI [...] will empower individuals to better \\ understand, interact with and contribute to the evolving landscape of AI, \\ fostering a more informed and participative society'' \citep[p. 6,][]{WEF_Doc3}.\end{tabular} \\ \hline
\begin{tabular}[c]{@{}l@{}}\textbf{Enabling Public} \\ \textbf{Scrutiny and} \\ \textbf{Monitoring}\\ \\ (encouraged by \\ n = 15 organisations)\\ \\ \vspace{35pt}\end{tabular} &
  \begin{tabular}[c]{@{}l@{}}\textbf{independent audits} of the system, e.g. through \\ members of the public or civil society organisations\\ \\ \textbf{organised post-deployment monitoring} via \\ regular system assessments, e.g. through collecting \\ feedback from operators and impacted non-users \vspace{55pt} \end{tabular} &
  \begin{tabular}[c]{@{}l@{}}SHI can be used in the examination of AI systems, from internal to fully inde-\\ pendent audits through the public or civil society. Such examination can vary \\ in its formality as well as its location in the AI lifecycle (very initial impact \\ assessments to post-deployment). A direct feedback loop with system \\ stakeholders enables them to report or alert harms, impacts, or changes in \\ the system's environment that require reactions. This allows the deployer \\ to make informed updates to the system, ensuring a lasting fit to its \\ evolving environment\vspace{35pt} \end{tabular} &
  \begin{tabular}[c]{@{}l@{}}HAI Stanford: ``[i]mplement executive and legislative actions to allow third-party auditor \\ access to AI data and source code, as well as other transparency and explainability \\ information, for the purposes of external researcher, civil society, and regulator \\ assessments'' \citep[p. 8,][]{Stanford_HAI}.\\ \\ EU High-Level Expert Group on AI: recommends to collect ``regular feedback even after \\ deployment'' \citep[p. 19,][]{HLEGAI_EC}. \\ \\ Access Now\citep[p. 20,][]{AccessNow_Doc}: ``[c]lear avenues should also \\ be established for people affected by AI systems, or groups representing them, to flag \\ harms and thereby trigger investigations by enforcement bodies'' \citep[p. 20,][]{AccessNow_Doc}.\end{tabular}
\end{tabular}%
}
\end{table}
\end{landscape}

\clearpage

\subsubsection{Who is Involved (and Who Not) Reflects Commercial Focus} 
\label{sec:SHI_scope}
\begin{figure*}[t!]
  \centering
  \includegraphics[width=.8\linewidth]{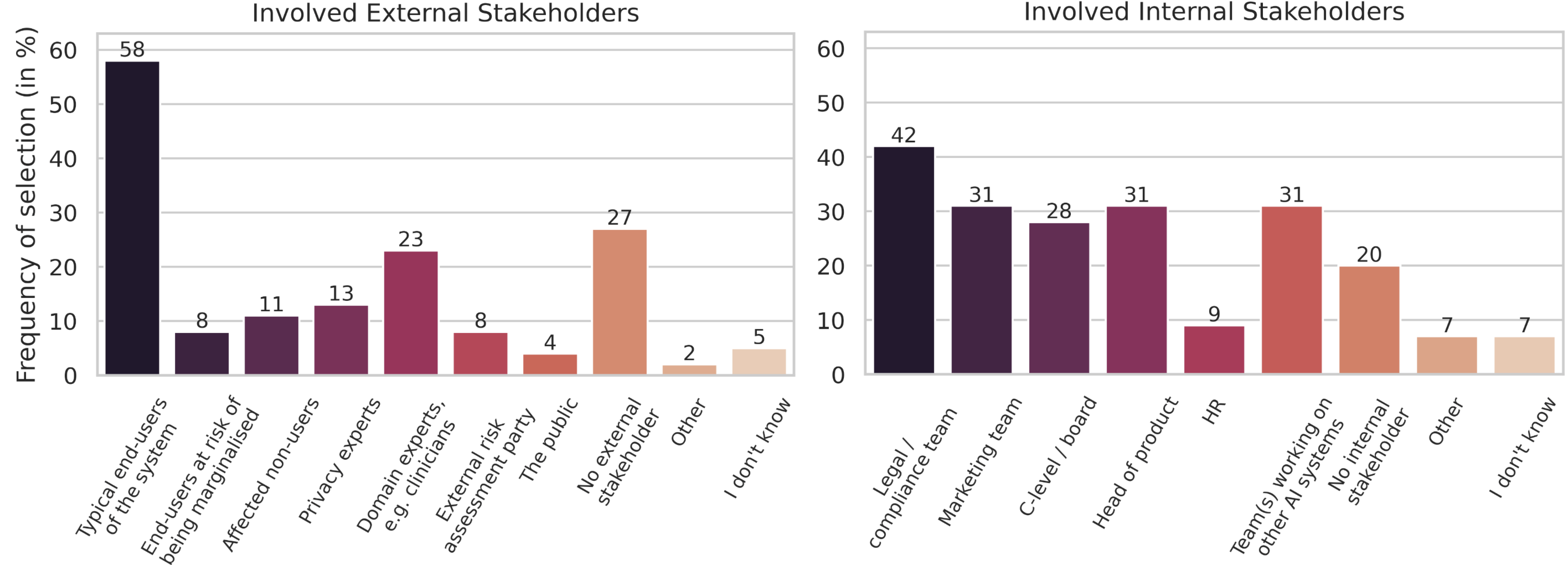}
  \caption{Frequency with which the survey participants indicated that a specific external or internal stakeholder was involved.}
  \label{fig:shs}
  \Description{Bar chart illustrating the proportions of survey participants that indicated that a specific stakeholder was involved (multiple choice possible).}
\end{figure*}

Next, we analyse which stakeholder groups were involved and which not, i.e. the scope of SHI. 
Our survey participants indicated that the two most engaged external stakeholder groups were typical end-users and domain experts (58\% and 23\%, Fig. \ref{fig:shs}). Since both groups can advance commercial interests through insights into the needs and expectations of users and customers, this suggests that \textbf{SHI involvement is heavily focused on such revenue-critical stakeholder groups}. Similarly, the internal legal team is the most involved internal stakeholder group, reflecting the importance of legal compliance (Fig. \ref{fig:shs}).
When asked who determines the scope of SHI, participants indicated the \textit{`developer team'} (44\%), the \textit{`C-suite'} (35\%), and the \textit{`internal legal team'} (35\%) as the main deciders whilst roles such as \textit{`ethics owner'} (14\%) or \textit{`UX researcher'} (9\%) had limited decision-agency. This finding reinforces the notion that commercial and compliance-related motivations are far more prevalent than contributing to rAI efforts.

Analysing who was \textbf{not} involved provides additional insights. Notably, only 4\% of survey participants indicated that they involved the general public, despite 28\% acknowledging that their system affects this group. Similarly, only 11\% involved affected non-users despite 32\% reporting them being affected by their system.
To investigate whether the anticipated impact of a system on a group influenced the involvement of that specific group, we conducted independent sample t-tests (\S\ref{sec:practitioner}, controlled for multiple comparisons). We performed separate tests for each stakeholder group (i.e. \textit{`user/operator'}, \textit{`impacted non-users'}, and \textit{`the public'}) to compare whether they were more\slash less involved when the system was (vs was not) expected to impact them. None of the tests showed significant differences, indicating that \textbf{affected communities that are not directly linked to commercial interests are rarely involved} in the design of systems, \textit{even if} they are expected to be impacted.

\subsubsection{Methods \& Timings of SHI Fail to Empower Stakeholders to Substantially Shape the System} 
\label{sec:SHI_methods}
\begin{figure*}[h]
  \centering
  \includegraphics[width=.7\linewidth]{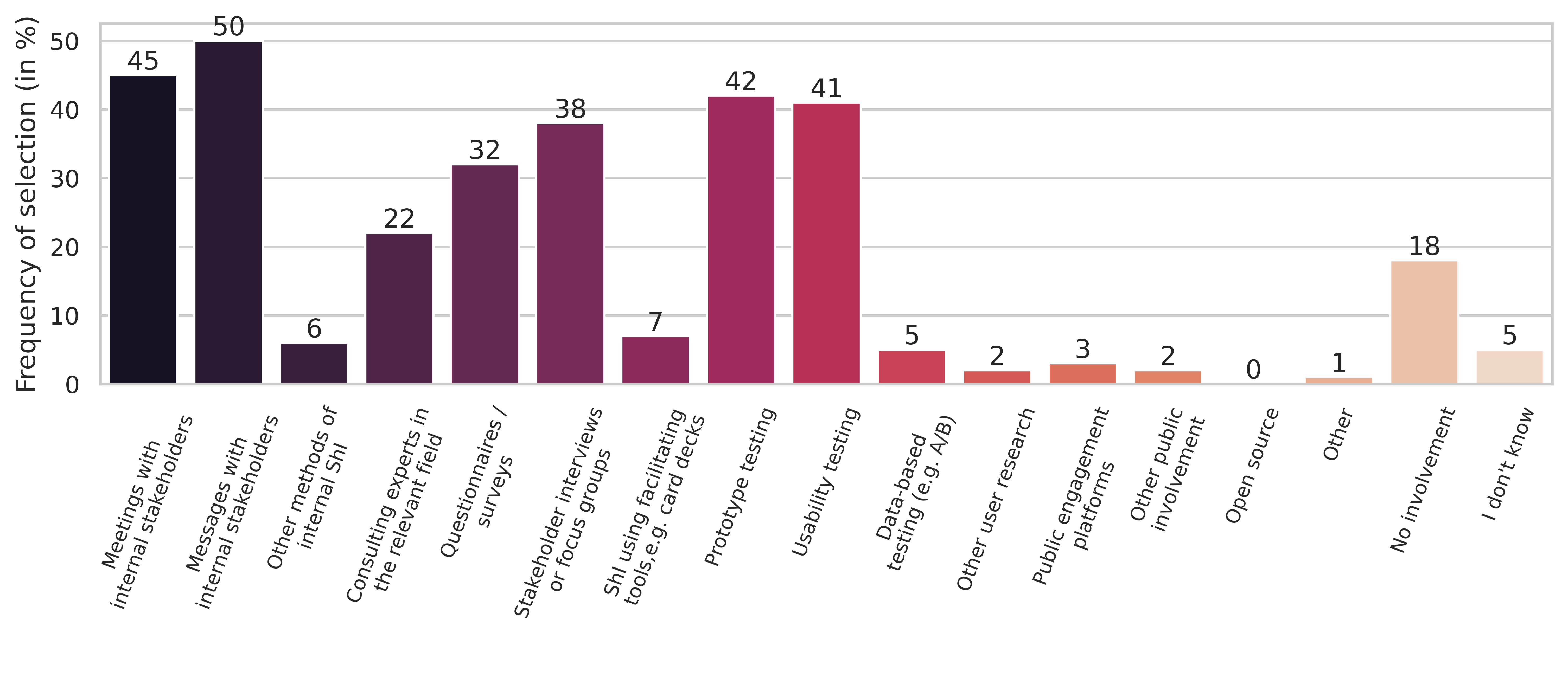}
  \caption{Methods of SHI as indicated by our survey participants.}
  \label{fig:methods}
  \Description{Bar chart illustrating the methods with which our practitioner sample reported to conduct SHI.}
\end{figure*}

Similarly, the \textbf{methods and timelines of SHI reflect commercial interests}, i.e. the traditional focus of SHI. External stakeholders were primarily engaged through prototype (42\%) or usability testing (41\%, Fig. \ref{fig:methods}). These methods are usually deployed at later stages where the main objective and features of the system have already been determined. 80\% of the interviewees confirmed that most external SHI efforts were concentrated at later stages (i.e. ``as soon as [...] we have something that we can give away or show'', I7). Even when external stakeholders were involved earlier, their involvement was limited to a few purely consultative engagements with a smaller sample size, rather than a more comprehensive and participatory approach. This is concerning, as SHI at the outset of development is crucial to align the system's purpose with the needs and values of the communities it will affect \cite{ISO2, corbett2023power}.

Notably, \textbf{none of the interviewees reported a collaborative identification of their system's objectives} with the intended users or affected non-users.
Instead, nearly all interviewees described a top-down process where the objective was imposed by their management\slash business strategy department or the management of a customer organisation. Often, the interviewees perceived this process as unclear and lacking transparency.
Only three interviewees mentioned to use external SHI to validate the top-down assigned objectives, i.e. gather SHI insights to determine whether it should actually be pursued.
This suggests that external stakeholders are primarily involved at stages where only smaller-scale changes (e.g. to the interface) are possible, while more impactful decisions are already decided upon and thus off-limits, e.g. determining an objective or whether AI is a valuable contribution to the solution. 
Open text survey responses suggest that some practitioners are aware and disapproving of this practice, e.g. expressed the wish to \textit{``involve the stakeholders from the early phases, cooperating with them towards an objective''}. Further, only 5\% of practitioners indicated methods that involve the public in any way and merely 7\% used facilitating tools during SHI.

\subsubsection{Commercial \& Internal Agendas Discourage SHI Beyond its Traditional Form} \label{sec:SHI_conflictcommercial}
Our analysis revealed that current SHI practices reflect traditional goals and methods of SHI (\S\ref{sec:SHI_drivers}, \ref{sec:SHI_scope}, \ref{sec:SHI_methods}). Critically, we further found that this discourages more comprehensive SHI practices: 80\% of our interviewees reported that their \textbf{organisations' commercial interests have potential conflicts with outputs\slash learnings resulting from SHI activities}. Interestingly, nearly all interviewees positioned themselves as mediators between commercial\slash internal and external stakeholder interests (potentially reflecting our SHI-attuned sample, \S\ref{sec:limitations}), trying to push for the latter whilst staying in the constraints of the former, i.e. ``hav[ing] to strike a balance between what the [internal] stakeholders want from us [...] versus what the end users expect'' (I5). 
Notably, nearly half of the interviewees expressed the wish that their management would prioritise broader SHI, i.e. that \textit{``they cared more about people than about money''} and 23\% of the survey participants named the \textit{company's attitude towards SHI} as a barrier to SHI. Confirming this, I4 gave an account where they identified significant risks for a vulnerable stakeholder group, only to be met with a management response focused solely on reputational (and thus commercial) harm, i.e., "when it's bad press and it's gonna affect our numbers, we'll do a bit" (I4).

In addition to conflicts with commercial interests, 70\% of our interviewees reported that \textbf{SHI insights might be in conflict with their management's or colleagues' personal beliefs} about desirable system objectives or features. These beliefs often seem motivated by professional agendas rather than of SHI insights: ``people are managing egos and careers’’ (I1). Since SHI might produce insights contradicting these preconceived notions (``we're going to get answers we don't want"), it ``can be a problem because people don't like criticism'' (I4). I4 further reported pressure to get SHI insights in line with his management's goals, i.e. ``try and get the answers that we want''. As a result, actual SHI is often discouraged, particularly for more fundamental decisions early in the development process.

These conflicts with internal, professional, or commercial agendas not only hinder the buy-in to conduct SHI but also impede the implementation of SHI insights \textit{even after they have been obtained}: nearly half of our interviewees reported the key challenge ``to make them stick to what you think would be good for the stakeholder'' (I6) or that ``[t]he hard part is [...] convincing people that it's important enough'' (I7). 
Taken together, these tensions between the traditional, commercial focus of SHI and SHI advancing rAI efforts likely explain the currently low priority of the latter.
The next section relates these findings to the benefits that rAI guidance associates with SHI for rAI development.

\subsection{Relating Practitioner Insights to rAI Guidance: Current SHI Fails to Advance rAI}
\label{sec:comparing}
We now explore how our insights into the SHI practice in industry align with the benefits associated with SHI in rAI guidance. This comparison clarifies the extent to which current SHI practices in industry already contribute to rAI efforts, the aim of this study. 
rAI guidance seems to promote SHI as an instrument to shift the focus and agency during AI development towards affected communities outside the organisations developing AI systems, i.e. to expand whose voice is considered worth hearing and reacting to (\S\ref{sec:benefits}). 
In contrast, our analysis of current SHI practices in industry revealed drivers resembling traditional aims of SHI, i.e. increasing customer value, usability, and thus commercial interests (\S\ref{sec:practitioner_practices}). As a result, the engagement of revenue-critical stakeholders and SHI for compliance matters are currently emphasised heavily while other impacted communities are comparatively neglected (\S\ref{sec:SHI_drivers}, \S\ref{sec:SHI_scope}). Notably, we identified a partially contradictory incentive structure between commercial interests and more responsible SHI practices (\S\ref{sec:SHI_conflictcommercial}).

Table \ref{tab:comparison} dives deeper into this disconnect. It contrasts each of the five benefits that rAI guidance associates with SHI with relevant findings from our practitioner studies. This mapping allows us to reflect on the extent with which currently established SHI practices are counteracting or contributing to the fulfilment of these five benefits. To summarise these reflections, we derive an overall estimate for each of the five benefits, i.e. estimate the extent to which currently established SHI practices can support the realisation of this benefit (\textit{italicised} line in cells describing current practices, Table \ref{tab:comparison}).

\begin{table*}[h]
\caption{Contrasting the five key benefits that rAI guidance hopes to advance through SHI (left col., \textbf{bold}) with our insights into current SHI practices (right col.). By mapping our industry insights to the five benefits we estimate the extent to which currently established practices can contribute to\slash contradict the achievement of each benefit (last line in the right col., \textit{italic})}
\label{tab:comparison}
\resizebox{\linewidth}{!}{%
\begin{tabular}{ll}
 \textbf{Benefits of SHI} \\ \textbf{for rAI (\S\ref{sec:benefits})} & \textbf{Our insights into the currently established SHI practice in AI industry (\S\ref{sec:practitioner_practices}) that contribute to\slash contradict the achievement of each benefit}\\
\hline
\begin{tabular}[c]{@{}l@{}}\textbf{Rebalance} \\ \textbf{Decision} \\ \textbf{Power}\end{tabular} &
  \begin{tabular}[c]{@{}l@{}}(1) whether a stakeholder was impacted by a system or not was independent from whether they were involved (\ref{sec:SHI_drivers})\\ 
  (2) stakeholders without direct commercial benefit (e.g. affected non-users or marginalised communities) are rarely included, reflecting the scope of traditional SHI (\ref{sec:SHI_scope})\\ 
  (3) the inclusion of voices of the wider public is not a common motivation for SHI (\ref{sec:SHI_drivers})\\ 
  (4) currently conducted SHI assigns only very limited agency to stakeholders, i.e. to assess rather than to shape a system (\ref{sec:SHI_methods})\\ 
  (5) commercial interests and internal agendas are prioritised over the needs of affected communities (\ref{sec:SHI_conflictcommercial}) \\
  $\rightarrow$ \textit{Estimated Contribution of Current Practices: \textbf{Very Low}}\end{tabular} \\ \hline
\begin{tabular}[c]{@{}l@{}}\textbf{Detailed} \\ \textbf{Understanding}\\ \textbf{of the Socio-} \\ \textbf{Technical}\\ \textbf{Context} \end{tabular} &
  \begin{tabular}[c]{@{}l@{}}(1) while SHI is used to understand needs and contexts, this is limited to stakeholders relevant to commercial interests (instead of the most affected or at-risk communities), \\ thus reflecting the traditional, commercially-motivated role of SHI in software development (\ref{sec:SHI_conflictcommercial}, \ref{sec:background})\\ 
  (2) commercial interests are dominant drivers of SHI, implying that e.g. affected non-users are out of scope, further limiting whose experience is understood \& included (\ref{sec:SHI_scope})\\ 
  (3) commercial\slash internal agendas are prioritised over addressing user needs, implying that they might be ignored even if identified (\ref{sec:SHI_conflictcommercial}) \\
  $\rightarrow$ \textit{Estimated Contribution of Current Practices: \textbf{Medium With Limited Scope}}\end{tabular} \\ \hline
\begin{tabular}[c]{@{}l@{}}\textbf{Improved Risk}\\ \textbf{Anticipation}\end{tabular} &
  \begin{tabular}[c]{@{}l@{}}(1) SHI is used to reduce harms, however, with a focus on legal, financial, or reputational harms only (e.g. via low customer value\slash satisfaction). \\ Harms without financial impact are neglected (\ref{sec:SHI_drivers}, \ref{sec:SHI_conflictcommercial})\\ 
  (2) identified harms\slash needs can be in conflict with commercial interests, thus impeding or even preventing reactions, aggravated by low stakeholder empowerment (\ref{sec:SHI_methods}) \\ 
  (3) narrow scope of involved stakeholders (vs involving the most affected or at-risk stakeholder groups) limits the risks that can be anticipated (i.e. risks to who) (\ref{sec:SHI_scope}) \\
  $\rightarrow$ \textit{Estimated Contribution of Current Practices: \textbf{Medium With Limited Scope}} \end{tabular} \\ \hline
\begin{tabular}[c]{@{}l@{}}\textbf{Increased Public}\\ \textbf{Understanding}\\ \textbf{and Trust}\end{tabular} &
  \begin{tabular}[c]{@{}l@{}}(1) since there is no direct commercial benefit \& insights might even be in conflict with such, the public is very rarely involved (\ref{sec:SHI_scope}, \ref{sec:SHI_conflictcommercial})\\ (2) increasing trust and inclusion of the wider public is not a common motivation for SHI (\ref{sec:SHI_drivers})\\ (3) narrow scope of SHI instead of including the broader public (\ref{sec:SHI_scope}) \\
  $\rightarrow$ \textit{Estimated Contribution of Current Practices: \textbf{Very Low}}\end{tabular} \\ \hline
\begin{tabular}[c]{@{}l@{}}\textbf{Enabling Public}\\ \textbf{Scrutiny and}\\ \textbf{Monitoring}\end{tabular} &
  \begin{tabular}[c]{@{}l@{}}(1) expert assessments e.g. through involving privacy or legal experts are only conducted if compliance (and thus commercial interests) are a concern (\ref{sec:SHI_drivers}, \ref{sec:SHI_scope}) \\ (2) conflicts of stakeholder insights with commercial interests render deliberate, public assessments an unlikely goal (\ref{sec:SHI_conflictcommercial})\\ (3) public involvement is rare and not a common driver behind SHI, impeding possibilities for public scrutiny (\ref{sec:SHI_drivers}, \ref{sec:SHI_conflictcommercial}, \ref{sec:SHI_scope})\\ (4) existing issues with conflicting insights and agendas suggest that further assessments and opinions might not be desired (\ref{sec:SHI_conflictcommercial}) \\
  $\rightarrow$ \textit{Estimated Contribution of Current Practices: \textbf{Low With Limited Scope}} \end{tabular} \\ \hline
\end{tabular}%
}
\end{table*}

Our analysis suggests that \textbf{current SHI practices in industry are not able to contribute to rAI efforts}, i.e. the benefits that rAI guidance associates with SHI are largely beyond what current SHI practices can achieve. Specifically, the benefits of \textit{rebalancing decision power} and \textit{increasing public participation and trust} are not considered as they fall outside the scope of traditional, commercially-motivated SHI.
A \textit{more detailed understanding of the socio-technical context} and \textit{improved risk anticipation} have potential for commercial gains, e.g. by reducing legal risks or clarifying customer expectations. Thus, current SHI practices are designed to realise these benefits for the subset of stakeholders that is directly tied to specific commercial interests. Since the resulting systems will provide more value and work better\slash more safely for those stakeholders than for those currently rarely involved (e.g. marginalised communities, impacted non-users), the latter are disadvantaged, thus contradicting equity-related goals of rAI efforts.
The benefit of \textit{enabling public scrutiny} is also limited: scrutiny is only enabled in relation to compliance matters, i.e. to avoid legal penalties by involving stakeholders such as auditors.
Therefore, \textbf{the current SHI focus on commercial incentives does not appear well-aligned with rAI efforts}. This accords with a theory from business ethics that distinguishes between SHI advancing ethical goals and SHI independent from these, depending on the level stakeholder involvement and decision-agency \cite{greenwood2007stakeholder}. 
Moreover, the conflicting incentives between commercial SHI practices and SHI supporting rAI efforts (\S\ref{sec:SHI_conflictcommercial}) highlight \textbf{tensions within existing power dynamics}, where the needs of marginalised communities are in conflict with the interests of AI companies \citep[see also][]{chan2021limits}. Given that AI companies currently hold most of the decision-making power, it seems unlikely that the practice will shift by itself towards more responsible SHI practices \citep[in line with previous works of the rAI community, e.g.][]{kallina2024stakeholder, kallina2025mapping, sloane2022participation, agnew2024inclusion, corbett2023power, rakova2021responsible}. Therefore, we need additional support to expand its scope, including interventions and improved incentive structures, e.g. through external pressure and commercial incentives. \S\ref{sec:recommendations} outlines potential avenues to realise this.

\section{Limitations}\label{sec:limitations}
When selecting rAI guidance documents, our main goal was to include a diverse range of organisations to identify shared reasons for recommending SHI for rAI development. However, we acknowledge representativeness limitations, particularly regarding geographically limited governmental rAI guidance, due to difficulties with retrieving translated (i.e. English) documents from countries with other languages. Despite this, we expect good coverage of main concepts regarding SHI benefits due to: (1) overlap and general agreement in government-issued rAI guidance content \citep[thus minimising the need to include an abundance of documents][]{jobin2019global, fjeld2020principled}), (2) English language dominance both in AI industry as well as AI governance and standards, and (3) highly reoccurring themes across analysed documents, suggesting good coverage. %
Further, we excluded rAI principles or guidance from the private sector to not conflate the goals of non-commercial rAI actors with the commercial, private sector practice. Future studies could investigate whether private sector rAI guidance associates the same benefits with SHI as non-commercial rAI guidance or whether it is shifted towards commercial interests.

We note that our participants may be skewed towards those interested in SHI as they were recruited using the term. However, since SHI-attuned practitioners seem more likely to seek and implement rAI guidance, such sample seems especially useful for exploring bottlenecks of SHI practices and their alignment with rAI guidance. Indeed, our findings already show a lack of SHI practices that contribute to rAI efforts; engaging practitioners less familiar with SHI may reveal even weaker or less aligned practices, highlighting the urgency of addressing this gap.
The number of interview participants (n = 10) reflects the difficulty of engaging expert AI practitioners -- as is required for gathering insights into current industry practices. Recruiting such sample is challenging regarding numbers and access, and especially so for studies like ours that require longer time commitments (interviews were between 56--81 minutes) to deep-dive into issues. Further, the interviews served as an elaboration on the challenges and patterns previously identified by our large survey sample (n = 130), thus designed to enrich existing findings instead of identifying further patterns. 
Whilst our expert sample enabled us to gather rich insights into the status quo, the demographic scope was relatively narrow: predominantly male and UK-based. This (unfortunately) appears to reflect broader trends in the AI practitioner workforce (over-proportionately male and Western \cite{AIIndex}); however, we encourage future work to expand beyond these demographics to support broader generalisability.

\section{A Path Forward}\label{sec:recommendations}
Building on our findings, we now outline key actions and research opportunities to advance SHI practices in line with the benefits identified in rAI guidance. While these suggestions are geared toward specific actors (such as academic researchers and regulator) and adjacent communities, they are likely to have broader relevance and should be seen as a starting point for further work.%

\subsection{Designing Clearer Guidance, Aiding Buy-In} \label{sec:rec_clearguidance}
Since this study showed that SHI can take various forms---many of which do not contribute to rAI---it is \textbf{essential that rAI guidance clearly states the early onset and continuity of the SHI process, the shift in decision-power, and the scope of involvement} required to advance rAI efforts. %
To move forward on this, we advocate for the co-creation of such guidance by rAI researchers, AI practitioners, and regulators to ensure that the resulting guidance is appropriate for different audiences and practitioner communities.
Importantly, this involvement should serve to add nuance and contextual awareness---for example, by highlighting external constraints such as regulatory or safety-critical requirements---rather than to dilute the guidance to better align with commercial (as opposed to broader societal) interests.
Additionally, it is essential that civil society organisations are given a meaningful voice and power, especially those representing at-risk communities. This will further ensure that the resulting guidance goes beyond traditional SHI and promotes SHI that contributes to rAI efforts \citep{Ada_Doc2}. 
Adding to the benefits of more concrete guidance, \citet{yildirim2023investigating} showed that rAI guidance documents can be instrumentalised as tools to obtain buy-in from the management\slash C-level: since they clearly set out the required steps for adhering to best practices, they help to communicate the need for concrete actions and their benefits towards deciders \cite{yildirim2023investigating}. Thus, we suggest further sharpening existing guidance so that employees can \textbf{demonstrate a clear and responsible SHI action plan to decision-makers}.

\subsection{More Clarity in Terminology} \label{sec:terminology}
As discussed in \S\ref{sec:comparing}, SHI has been part of software engineering for decades, but with a different focus than what recent rAI guidance is aiming to achieve. Our findings clearly show that currently common SHI practices must evolve to be able to contribute to rAI efforts. Additionally, our findings regarding the benefits that rAI guidance associates with SHI (\S\ref{sec:benefits}) suggest that SHI should not only occur during system development, but also as an ongoing process when testing, monitoring, or auditing a system, including post-deployment. 
To increase the awareness of these distinctions, \textbf{a more specific term for SHI that contributes to rAI efforts} seems crucial to avoid confusion and `ethics washing' \cite{wagner2018ethics}. In addition to further investigating and clarifying the different subtypes of SHI and their purpose, such clarity in terminology can help the development of associated best practices.
We suggest following distinctions as a starting point to further refine types of SHI: 

\small
\begin{itemize}
    \item \textbf{Participatory development} to refer to the involvement of affected communities in the system's development process \cite[e.g.][]{delgado2023participatory, corbett2023power}, but explicitly requiring that affected non-users \& at-risk communities are involved \citep[following][]{haque2024policing, chiu2024balancing, solyst2023youth}.
    \item \textbf{Public participation} to describe civil society involvement, either during AI development or the creation of AI or domain-specific regulation\slash policies \cite[e.g.][]{White_House_Terminology}.
    \item \textbf{Expert involvement} to refer to the consultation of internal or external domain experts not representing affected communities, e.g. human rights or social justice experts.
    \item \textbf{Public oversight mechanisms} to describe public participation in scrutiny \& AI policy enforcement, e.g. public audits \citep[e.g.][]{birhane2024ai, Yurrita_2022}.
\end{itemize}
\normalsize

\begin{table*}[h]
\caption{Starting point for future research, required to identify more concrete, actionable, \& rAI-aligned SHI guidance.}
\label{tab:recommendations}
\resizebox{\linewidth}{!}{%
\begin{tabular}{l}
\hline
\textbf{IMMEDIATE RESEARCH AGENDA} \\ \hline

\textbf{Understanding SHI along the AI lifecycle and across use cases:}\\ \hline
(1) How should SHI manifest at the various stages of the AI lifecycle and what is the aim per stage? How do SHI best practices change for the different stages of the AI lifecycle? \\
(2) It is rarely feasible to engage ``everyone who is impacted'' (\S\ref{sec:intro}). Which factors of the use case make SHI more (or less) pressing? \\
(3) Which factors determine the scope of involvement (i.e. which communities should be involved) and beneficial SHI methods? \\
(4) What can we learn from SHI in non-AI contexts? How does this change with foundation models where use cases are unclear? \\ 
(5) For use cases in which the software provider is detached from the affected communities (e.g. third-party suppliers), how can we ensure the involvement of downstream stakeholders? \\ \hline

\textbf{Designing improved SHI guidance and tools:} \\ \hline
(1) What is the ideal balance between universal and implementation-specific guidance? How do other industries handle this trade-off? How should more universal guidance be translated to specific use cases? \\
(2) Only 7\% of our practitioners used supporting tools during SHI. How can existing HCI methods \cite[e.g.][]{tosi2020user, Interaction_Design} help to create more actionable tools?\\
(3) How can tensions between conflicting stakeholder needs or between stakeholder needs and commercial interests be addressed? \\ 
(4) Why and how do different practitioner roles conduct SHI (e.g. UX, customer support, ethics owner, developer) and how can we support them better? \\ \hline

\textbf{Informing AI regulation \& public participation and engagement:} \\ \hline
(1) Which aspects of AI regulation can effectively incentivise SHI in line with rAI goals, beyond recommendations in \S\ref{sec:regulationlever}?\\
(2) How can we ensure that regulation incentivises SHI with specific stakeholders (especially communities outside of commercial motivations) and with substantial decision-agency?\\ 
(3) What are accessible and respectful avenues for public participation and how can the public be incentivised and rewarded for engagement? \\
(4) How can we facilitate the access to at-risk communities and reward representatives for their commitments? What can we learn from related, not AI-specific fields such as anthropology and social justice? \\
(5) How can conflicting incentives in industry (i.e. commercial interests vs needs of at-risk groups) be resolved or mitigated? \\ \hline
\end{tabular}
}
\end{table*}

\subsection{Regulation: A Lever for Aligning Incentives} \label{sec:regulationlever}
Our insights suggest that legal pressure could be a significant lever in driving the adoption of more rAI-oriented SHI. This is in line with a report that demonstrated that industry practitioners prioritise rAI principles related to compliance over broader rAI efforts \cite{EY_Report}. The findings of the report suggest that practitioners prioritise or even filter rAI efforts (such as rAI principles or SHI) by the extent to which they are required to comply with regulation -- and less on the specific impacts they might have on affected communities. Supporting this, tech companies were shown to associate the GDPR primarily with commercial and legal risks vs with ethical obligations \cite{wong2023privacy, norval2021data}. %
Therefore, \textbf{using law to tie rAI-advancing SHI more directly to commercial interests seems a powerful lever} to better align the commercial practice with rAI efforts, e.g. by using both hard and soft law to adjust current incentive structures so that they promote involvement beyond revenue-critical stakeholders. This finding is especially relevant in a time in which significant effort is put into policing and regulating AI systems. The rAI research community could add much value by exploring how SHI with different and especially marginalised stakeholder groups could be incentivised effectively.

Given the generally low level of SHI during defining system objectives---despite objectives having the most fundamental impact on the values embedded in the system \cite{corbett2023power, cobbe2021reviewability}---, we suggest \textbf{strong legal incentives for SHI from the very start of system development lifecycle and throughout}. These could draw from existing regulatory approaches, e.g. use impact assessments as already required by the GDPR and the EU AI Act \citep{janssen2022practical, EUAIAct}: a mandatory impact assessment at the very outset of the AI lifecycle could require reporting on how the system's objective was identified and agreed upon in collaboration with affected communities, especially emphasising how at-risk communities were empowered to shape this objective. 
Further, future standards on SHI for rAI efforts could become directly referenced in AI policy \cite[see][]{abbott2009politics, leibrock2002methods, casey2009three}, expanding beyond current links between rAI guidance and regulation (\S\ref{sec:AIpolicy}), thereby integrating the benefits of standards into policy (e.g. clarity, rapid implementation, regular revisions, and the ability to span jurisdictions \cite{ISOstand_months, renckens2020private}). Indeed, the  EU AI Act already references standards as important, ``value-adding services'' \cite[Chapter 6, Article 58,][] {EUAIAct}. Thus, we strongly encourage the rAI community to work on evolving and clarifying existing guidance to facilitate the emergence of rAI-aligned SHI standards. \S\ref{sec:researchagenda}, we outline key areas of investigation to support this goal.

\subsection{Immediate Research Agenda} \label{sec:researchagenda}
Our work clearly demonstrates a range of opportunities for further research that addresses the remaining concerns. This encompasses the characteristics of beneficial SHI practices and their interaction with specific use cases---such as what type and scope of SHI are most important or effective for a given system and when; all enabling more specific SHI guidance---as well as how the SHI process aligns with various stages of the AI development lifecycle, including the type and scope of SHI required at each stage.

As such, we now present an initial research agenda---outlined in Table \ref{tab:recommendations}---as a starting point for gathering the insights required for more concrete, actionable, and effective guidance concerning SHI contributing to rAI efforts. %

\section{Conclusion}
Our work reveals that established SHI practices in commercial AI settings are largely misaligned with rAI efforts. \textbf{While rAI guidance promotes SHI to address asymmetries in agency, priorities, and knowledge, the currently established SHI practices focus on advancing commercial interests}. Thus, current SHI is limited to revenue-critical stakeholders while affected and particularly at-risk communities are not considered or empowered to shape the systems.
Moreover, we found \textbf{conflicts between stakeholder needs and commercial interests} that are likely to discourage the shift towards more responsible SHI practices.
Clarifying the differences between established SHI practices and those aligned with rAI efforts is essential for helping rAI and adjacent communities respond in timely and informed ways. 

We hope the insights offered here will support collaboration across the rAI research community, guidance bodies, and practitioners -- particularly through the
\textit{following key actions}.
First, we wish to motivate the co-design of improved and actionable guidance on rAI-advancing SHI. This is required to derive standards, enable reflections in industry on established practices and improvements, as well as to obtain buy-in from deciders (\S\ref{sec:rec_clearguidance}). 
Second, we call for direct efforts of the academic rAI communities to address open research questions such as those proposed in \S\ref{sec:researchagenda}, designed to identify and enable beneficial practices in line with rAI efforts going forward.
Lastly, the revealed disconnects between rAI guidance and AI industry are invaluable insights for AI policy makers and regulators, allowing them to tailor regulatory efforts as well as broader interventions towards addressing these.
In all, our findings contribute to efforts of further aligning current practices in commercial AI settings with rAI efforts to ultimately enable more responsible technologies.

\bibliographystyle{ACM-Reference-Format}
\bibliography{sample-base}

\newpage
\appendix

\section{Appendix A - Overview Over Organisations Purporting rAI Guidance}
\label{appendix:orglist}
This section provides an overview over the methods we deployed to address Q1, i.e. to derive the benefits that rAI guidance associates with SHI. This includes details on how we compiled the list of organisations purporting rAI guidance, the process of retrieving relevant documents from these organisations, as well as the resulting list of organisations and the analysed documents

\subsubsection{Generating a List of Organisations Publishing rAI Guidance and Their Relevant Publications}
We compiled a list of organisations publishing rAI guidance given we did not discover peer-reviewed list of such. Our aim was to include a broad range of organisations as to draw out common reasons for recommending SHI for rAI development. We hope that this list of organisations issuing rAI guidance will be harnessed by future studies analysing rAI guidance.
To create the list, we used prior mappings of organisations in the rAI space \citep{OECD_List, T20_List} and added organisations based on convenience sampling \citep{sedgwick2013convenience}. This first iteration was extended through input from four experts in complementary fields, each active for at least four years in their fields, i.e. one expert per AI governance, civil society action, technical approaches to rAI, and Human-Computer Interaction. This resulted in a list with 29 organisations, spanning multi-stakeholder organisations, inter-governmental and governmental organisations, as well as civil society.
For each of the organisations on this list, we reviewed all available rAI guidance publications on their websites and retrieved 56 guidance documents or recommendations regarding SHI during the development or governance of AI systems. The organisations and guidance documents are listed in Table \ref{tab:org}. Note that several of the included documents had input from private sector employees, however, we note the difference between documents developed and published by the private sector and those published by other organisations incorporating such input.

\begin{table*}[]
\caption{Overview over the organisations and their published the rAI guidance documents that were included in our thematic analysis (\S\ref{sec:AIpolicy}) to identify the beneficial outcomes that rAI guidance associates with SHI for rAI development}
\label{tab:org}
\resizebox{\linewidth}{!}{%
\begin{tabular}{ll}
\hline
\textit{\textbf{ORGANISATION}} & \textit{\textbf{ANALYSED RAI GUIDANCE DOCUMENT (TITLES AND YEAR)}} \\ \hline
\multicolumn{2}{l}{\textbf{Multi-stakeholder Organisations}} \\ \hline
\multirow{2}{*}{\textit{ACM - Technology Policy Council}} & - Statement on Principles for Algorithmic Systems (2022) \citep{ACM_Doc_2} \\
 & - Principles for the Development, Deployment, and Use of Generative AI Technologies (2023) \citep{ACM_Doc_1} \\ \hline
\textit{Data \&amp; Society} & - Democratizing AI: Principles for Meaningful Public Participation (2023) \citep{DemocratisingAI} \\ \hline
\textit{IEEE} & - Ethically Aligned Design (2019) \citep{IEEE_Doc} \\ \hline
\textit{University of Montreal} & - Montreal Declaration for a Responsible Development of AI (2018) \citep{Montreal} \\ \hline
\multirow{2}{*}{\textit{Partnership on AI}} & - Guidelines for AI and Shared Prosperity (2023) \citep{PartnerSHIpAI_Doc1} \\
 & - Making AI Inclusive: 4 Guiding Principles for Ethical Engagement (2022) \citep{PartnerSHIpAI_Doc2} \\ \hline
\multirow{3}{*}{\textit{The Alan Turing Institute}} & - Understanding AI Risks and Safety (2019) \citep{Turing_Doc} \\
 & - Course: AI Ethics and Governance (2022) \citep{Turing_Course} \\
 & - Data Justice in Practice (2022) \citep{Turing_Doc3} \\ \hline
\textit{HAI Stanford} & - Stanford HAI Artificial Intelligence Bill of Rights (2022) \citep{Stanford_HAI} \\ \hline
\textit{Beijing Academy of AI} & - Beijing AI Principles (2019) \citep{Beijing_Doc} \\ \hline
\textit{AI Now Institute} & - Advancing Racial Equity Through Technology Policy (2023) \citep{AI_NowDoc1} \\ \hline
\multirow{2}{*}{\textit{Ada Lovelace Institute}} & - Rethinking Data and Rebalancing Digital Power (2022) \citep{Ada_Doc1} \\
 & - Inclusive AI Governance (2023) \citep{Ada_Doc2} \\ \hline
\textit{Future of Life Institute} & - Position Paper on the EU AI Act (2023) \citep{FutureOL_Doc} \\ \hline
\multicolumn{2}{l}{\textbf{\textbf{Intergovernmental Organisations}}} \\ \hline
\multirow{3}{*}{\textit{UNESCO}} & - Recommendation on the Ethics of Artificial Intelligence (2021) \citep{UNESCO_Doc} \\
 & - Ethical Impact Assessment: A Tool of the Recommendation on the Ethics of Artificial Intelligence (2023) \citep{UNESCO_Doc2} \\
 & - Keyfacts: UNESCO’s Recommendation on the Ethics of Artificial Intelligence (2023) \citep{UNESCO_Doc3} \\ \hline
\multirow{2}{*}{\textit{UN}} & - A United Nations System-Wide Strategic Approach and Road Map for Supporting Capacity Development on AI (2019) \cite{UN_Doc} \\
 & - Principles for the Ethical Use of Artificial Intelligence in the United Nations System (2022) \cite{UN_Doc2} \\ \hline
\multirow{3}{*}{\textit{OECD}} & - OECD AI Principles (2019) \citep{OECD_Doc} \\
 & - Legal Instruments Artificial Intelligence (2019, last update 2023) \citep{OECD_Legal} \\
 & - AI in Society - Public Policy Considerations (2019) \citep{OECD_PublicPolicy} \\ \hline
\multirow{2}{*}{\textit{GPAI}} & - Responsible AI Working Group Report (2023) \citep{GPAI_Doc1} \\
 & - Scaling Responsible AI Solutions (2023) \citep{GPAI_Doc2} \\ \hline
\multirow{4}{*}{\textit{Council of Europe}} & \begin{tabular}[c]{@{}l@{}}- Recommendation of the Committee of Ministers to Member States on the Human Rights Impacts of Algorithmic \\ Systems (2020) \citep{CouncilofEurope_Doc1}\end{tabular} \\
 & \begin{tabular}[c]{@{}l@{}}- Possible Elements of a Legal Framework on Artificial Intelligence, Based on the Council of Europe’s Standards \\ on Human Rights, Democracy and the Rule of Law (2021) \citep{CouncilofEurope_Doc2}\end{tabular} \\
 & - Resolution: Need for Democratic Governance of Artificial Intelligence (2020) \citep{CouncilofEurope_Doc3} \\
 & - Unboxing Artificial Intelligence: 10 Steps to Protect Human Rights (2019) \citep{CouncilofEurope_Doc4} \\ \hline
\textit{G20 / T20} & - Why the G20 Should Lead Multilateral Reform for Inclusive Responsible AI Governance for the Global South (2023) \citep{T20_List} \\ \hline
\multirow{3}{*}{\textit{WEF}} & - The Presidio Recommendations on Responsible Generative AI (2023) \citep{WEF_Doc1} \\
 & - We Must Come Together to Ensure an AI Future That Works for All (2023) \citep{WEF_Doc2} \\
 & - Nine Ethical AI Principles for Organisations to Follow (2021) \citep{WEF_Doc3} \\ \hline
\textit{WHO} & - Ethics and Governance of Artificial Intelligence for Health (2021) \citep{WHO_Doc} \\ \hline
\multicolumn{2}{l}{\textbf{Civil Society Organisations}} \\ \hline
\multirow{3}{*}{\textit{Access Now}} & - Human Rights Impact Assessments for AI: Analysis and Recommendations (2022) \citep{AccessNow_Doc} \\
 & - Tech and Conflict: a Guide for Responsible Business Conduct (2023) \citep{AccessNow_Doc2} \\
 & - The European Human Rights Agenda for the Digital Age (2019) \citep{AccessNow_Doc3} \\ \hline
\multirow{3}{*}{\textit{Amnesty International}} & - Global: Companies Must Act Now to Ensure Responsible Development of Artificial Intelligence (2023) \citep{Amnesty_2} \\
 & - Ethical AI Principles Won’t Solve a Human Rights Crisis (2019) \citep{Amnesty_Doc1} \\
 & - Comment on Proposed Artificial Intelligence Regulations (2020) \citep{Amnesty_Doc3} \\ \hline
\multirow{2}{*}{\textit{AlgorithmWatch}} & \begin{tabular}[c]{@{}l@{}}- On the European Commission’s White Paper on Artificial Intelligence – a European Approach to Excellence \\ and Trust (2020) \citep{AlgorithmWatch_Doc1}\end{tabular} \\
 & - Sustainable AI in Practice (2022) \citep{AlgorithmWatch_Doc2} \\ \hline
\multirow{2}{*}{\textit{Centre for Democracy \& Technology}} & - CDT, Civil Society Reps to UK AI Safety Summit Urge Focus on AI Risks to People’s Rights (2023) \citep{CDT_Doc} \\
 & - CDT and Public Interest Organizations Call on Congress to Consider Current Harms of AI (2023) \citep{CDT_Doc2} \\ \hline
\multirow{4}{*}{\begin{tabular}[c]{@{}l@{}}\textit{Shared Publications of Various}\\ \textit{Civil Society Organisations}\end{tabular}} & - The Toronto Declaration: Protecting the Rights to Equality and Non-Discrimination in ML Systems (2018) \citep{Toronto_Doc} \\
 & - EU Trilogues: The AI Act Must Protect People’s Rights (2023) \citep{Shared_Doc1} \\
 & - Negotiations of Council of Europe Convention on AI, Rule of Law, Human Rights and Democracy (2023) \citep{Shared_Doc2} \\
 & - An EU Artificial Intelligence Act for Fundamental Rights (2021) \citep{Shared_Doc3} \\ \hline
\multicolumn{2}{l}{\textbf{Selection of Governmental Organisations}} \\ \hline
\begin{tabular}[c]{@{}l@{}}\textit{Chinese National Governance}\\ \textit{Committee for AI}\end{tabular} & - Governance Principles for Responsible AI (2019) \citep{China_original, translation} \\ \hline
\multirow{2}{*}{\begin{tabular}[c]{@{}l@{}}\textit{EU - European High Level Expert}\\ \textit{Group on AI \&amp; European Commission}\end{tabular}} & - Ethics Guidelines for Trustworthy AI (2019) \citep{HLEGAI_EC} \\
 & - EU AI Act (2024) \citep{EUAIAct}\\ \hline
\textit{Smart Africa} & - AI for Africa Blueprint \citep{SmartAfrica_Doc} \\ \hline
\multirow{3}{*}{\begin{tabular}[c]{@{}l@{}}\textit{USA - White House Office of Science} \\ \textit{and Technology Policy \& NIST}\end{tabular}} & - Blueprint for an AI Bill of Rights (2022) \citep{US_Doc} \\
 & - Executive Order on the Safe, Secure, and Trustworthy Development and Use of Artificial Intelligence (2023) \citep{US_Doc2} \\
 & - Artificial Intelligence Risk Management Framework (AI RMF 1.0) (2023) \citep{US_Doc3} \\ \hline
\end{tabular}%
}
\end{table*}

\subsubsection{Analysis of Relevant Publications}
The relevant 56 publications were systematically analysed, using both skimming technique \cite{dhillon2020effect} as well as a set of keywords associated with SHI, thus likely to identify relevant paragraphs. The keywords were selected based on commonly used words in the context of SHI across regulatory and academic publications, i.e. : \textit{stakeholder}, \textit{consultation}, \textit{involvement} / \textit{involve}, \textit{engagement} / \textit{engage}, \textit{civil}, \textit{society}, \textit{participation} / \textit{participatory} / \textit{participate}, \textit{public}, \textit{community} / \textit{communal}, and \textit{affected}. On the relevant paragraphs, we performed a manual thematic analysis following \citet{braun2012thematic} using the software Atlas.ti \citep{Atlasti}. The analysis resulted in 5 themes and 15 sub-themes that can be viewed in \S\ref{sec:benefits}.

\clearpage

\onecolumn

\section{Appendix B - Sample Details}
\label{appendix:sample}

\subsection{Survey Sample}

The following section provides details regarding the demographics and employment (roles, experience, and employers) of the survey sample.

\begin{figure*}[h]
  \includegraphics[width=.7\linewidth]{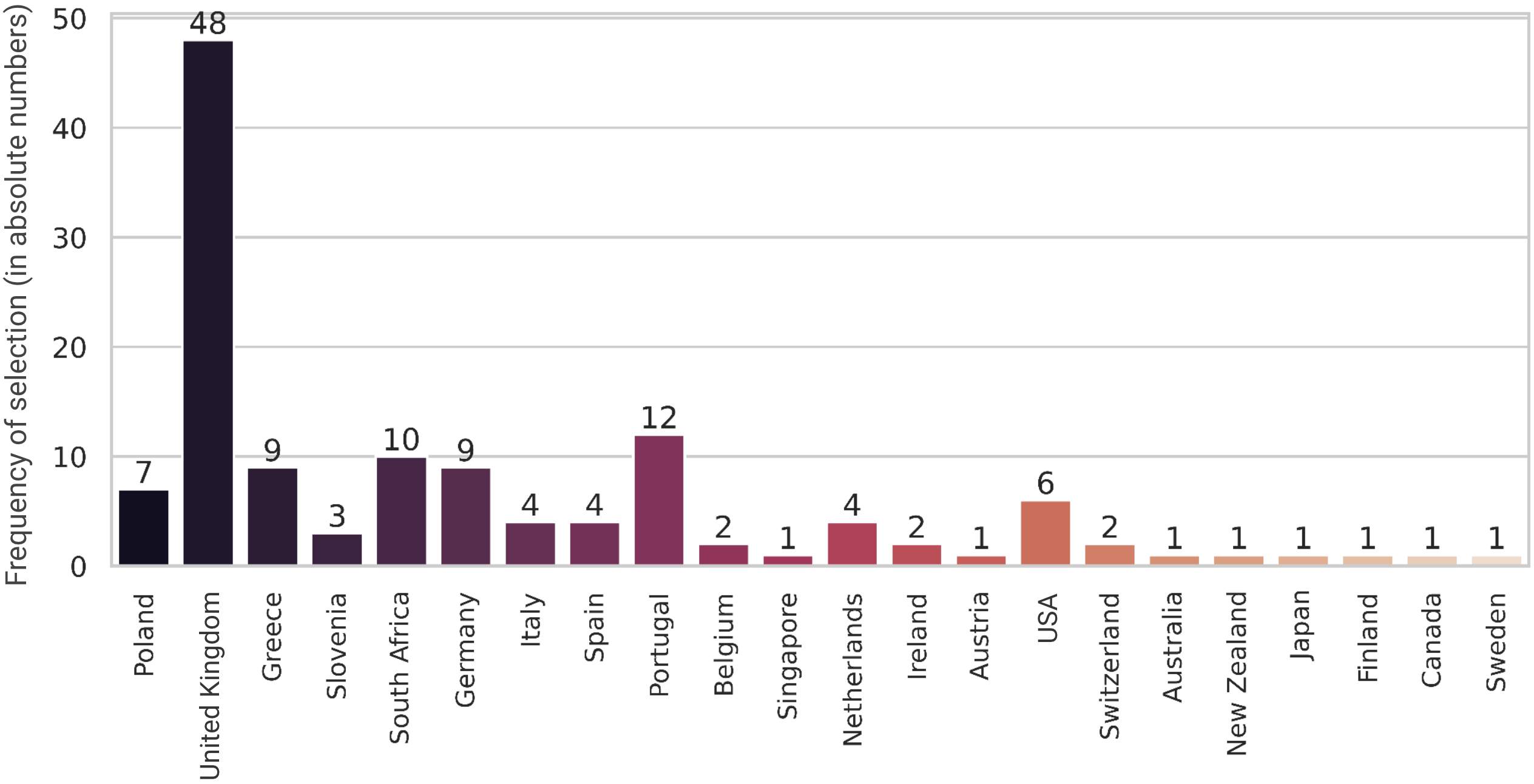}
  \caption{Country of residence as indicated by the survey participants (single-choice question).}
  \label{fig:origin}
  \Description{Bar chart illustrating the country of residence of the survey participants.}
\end{figure*}

\begin{figure*}[h]
 \centering
  \includegraphics[width=.7\linewidth]{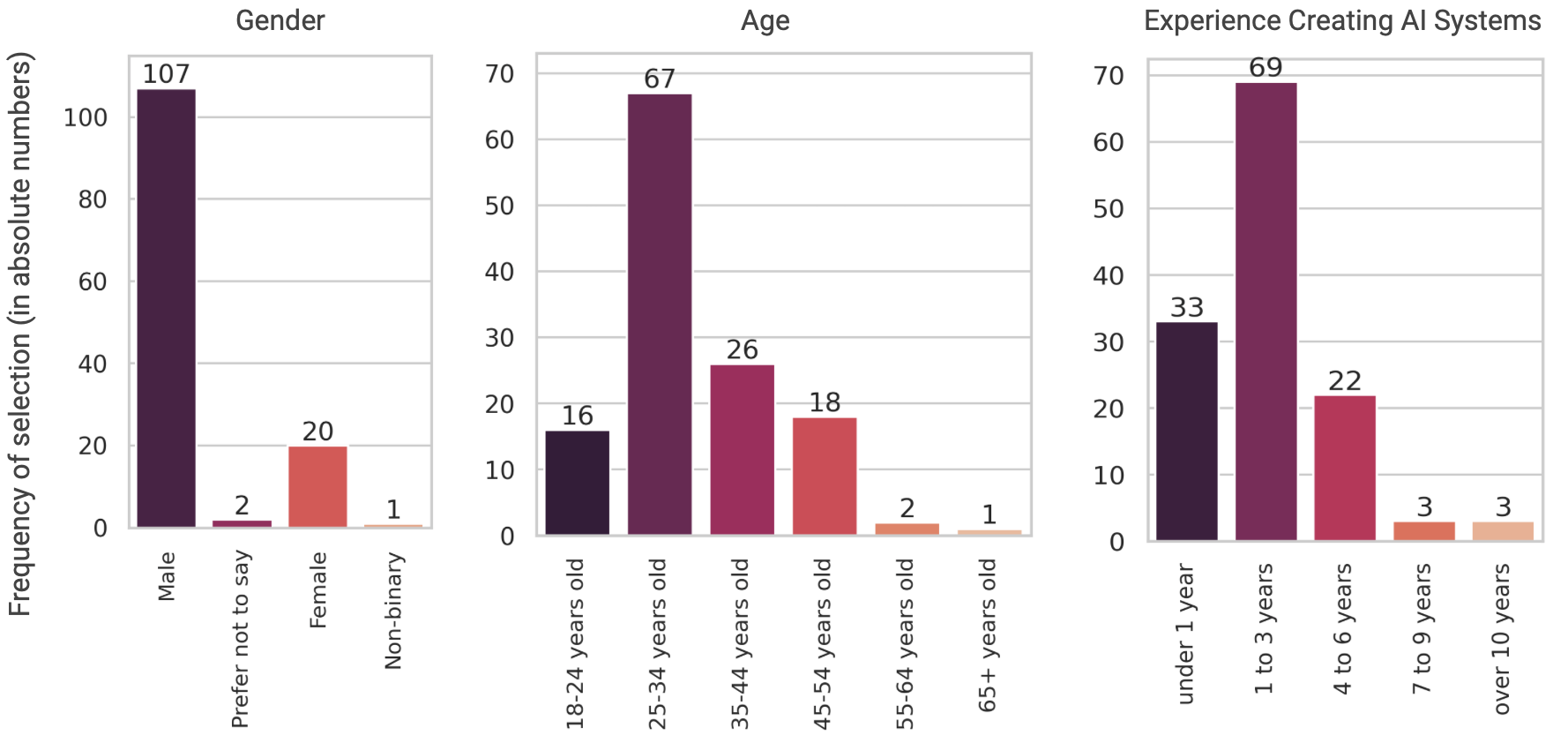}
  \caption{Gender and age distribution of the survey participants as well as the duration of their experience working along the creation of AI systems.}
  \label{fig:agegender}
  \Description{Bar chart illustrating the age, gender, and work experience of the survey participants.}
\end{figure*}

\textbf{Demographics.} As illustrated in Fig. \ref{fig:agegender}, 82\% of the \textbf{survey} participants indicated to be male and 15\% female (1\% non-binary, 2\% preferred not to say). The most common age was `25 to 34' (52\%; 20\% were `35 to 44' and 14\% `45 to 54'). Fig. \ref{fig:origin} illustrates that the majority of the survey participants lived in Europe (84\%), particularly the UK (37\%). 

\begin{figure*}[h]
  \includegraphics[width=\linewidth]{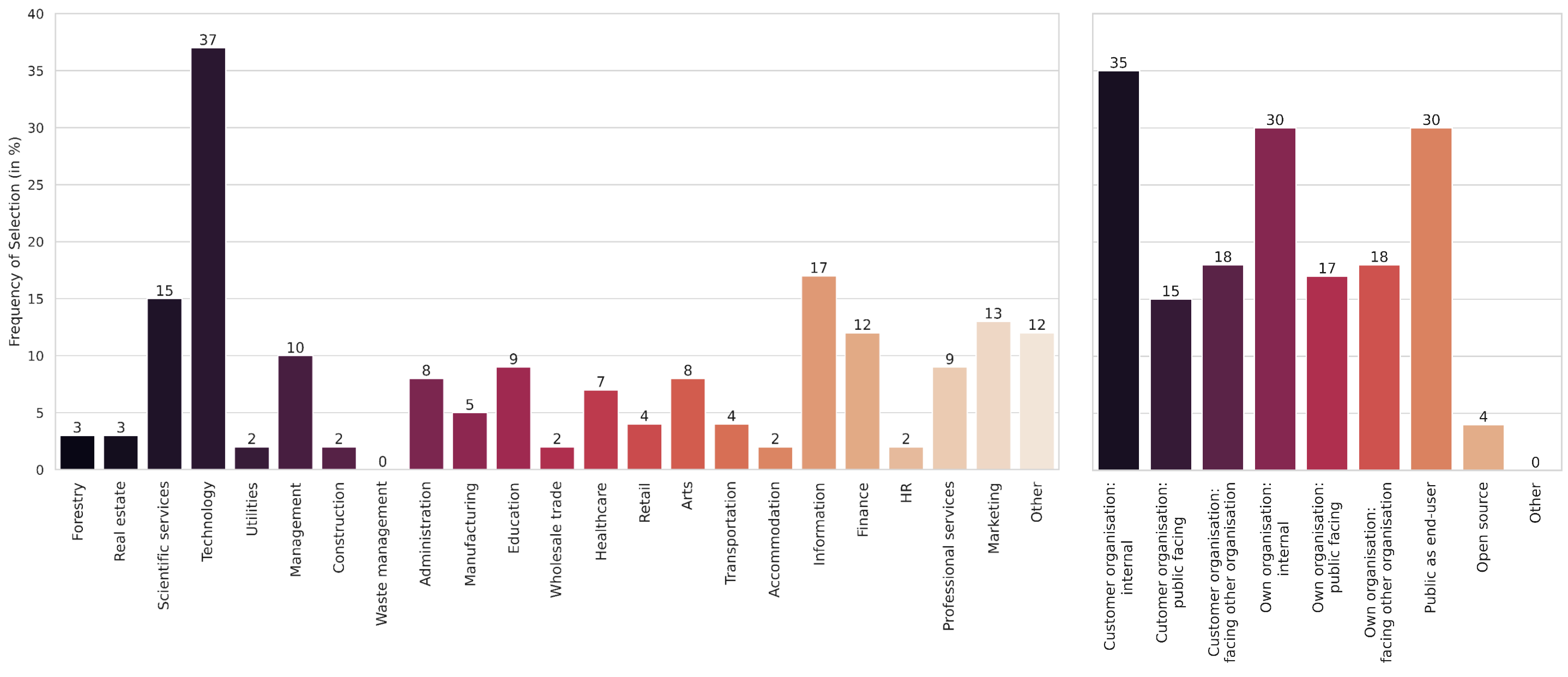}
  \caption{Industry for which the survey participants are creating AI systems as well as the planned type of deployment of these systems.}
  \label{fig:Industries}
  \Description{Bar chart illustrating the industry for which the survey participants are creating AI systems as well as the planned type of deployment of these systems.}
\end{figure*}

\begin{figure*}[h]
  \centering
  \includegraphics[width=\linewidth]{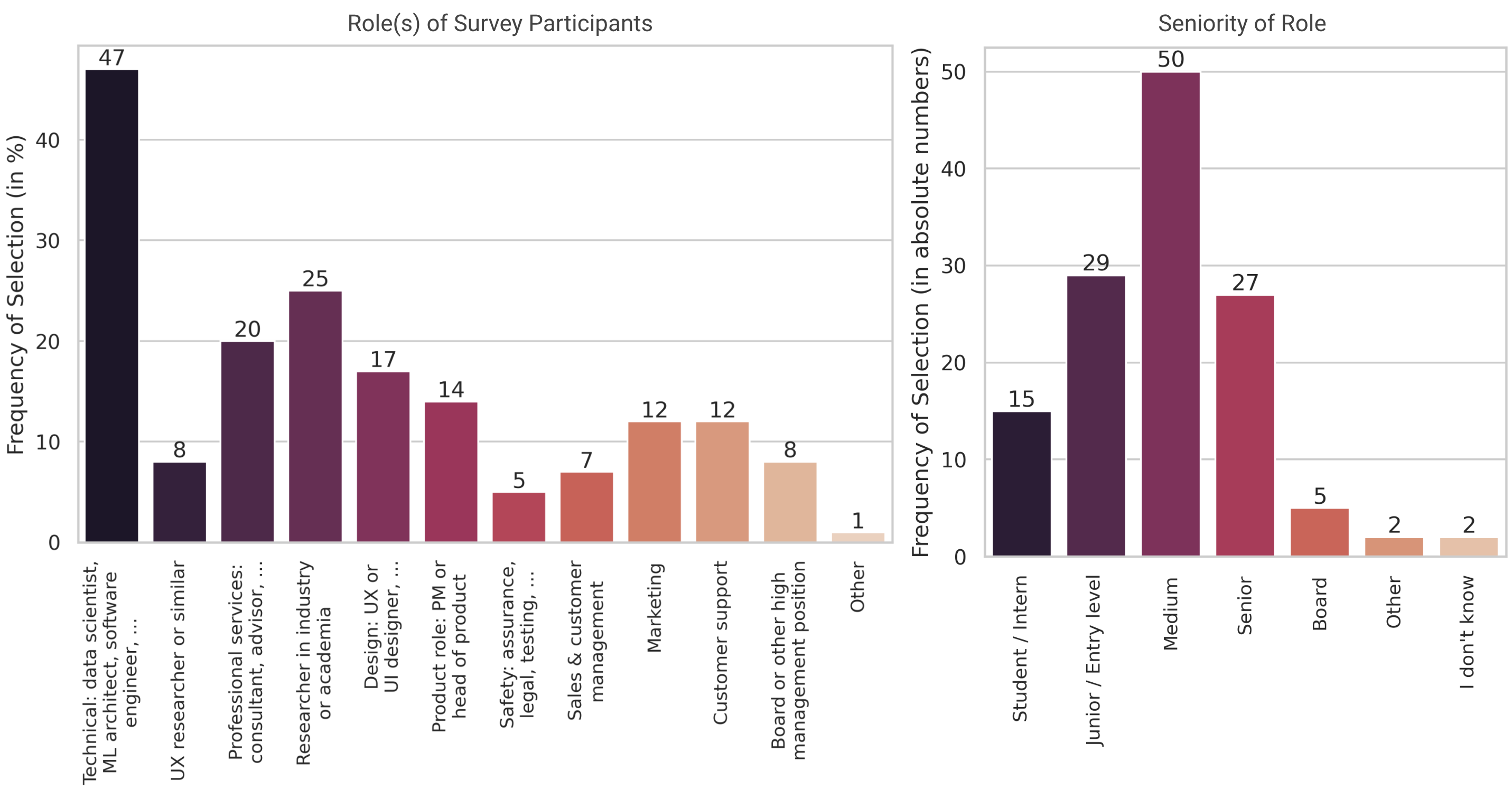}
  \caption{Roles in which the survey participants worked along the creation of AI systems (multiple-choice) as well as the seniority of this role.}
  \label{fig:roles}
  \Description{Bar chart illustrating the roles of the survey participants.}
\end{figure*}

\textbf{Employment.} The survey participants answered most questions for one specific system they work(ed) on in their career. According to the participants, the most prominent industries of these systems were Technology (37\%), Information sector (17\%), and Scientific Services (15\%) (Fig. \ref{fig:Industries}). Most systems were for internal use, either for a customer organisation (35\%) or the organisation employing the participant (30\%). 30\% were for customer-facing use whereby the customer was the general public.
53\% of our survey participants worked `1 to 3 years' along the creation of AI systems (25\% under `under 1 year' and 17\% `4 to 6 years') and 38\% indicated that their role had medium seniority (Fig. \ref{fig:agegender} and Fig. \ref{fig:roles}). 
Most participants indicated to work in technical roles (47\%), as researchers (academia or industry, 25\%), as consultants or advisors (20\%), designers (17\%), or in product roles (14\%) (multiple choice possible, Fig. \ref{fig:Industries}).

\clearpage

\subsection{Interview Sample}
Details on those participating in the \textbf{interviews} can be viewed in Table \ref{tab:interviewees} (row number corresponds to interviewee number in the results section). Our sample size was restricted by the number of survey participants that were willing to participate in the follow-up interviews as well as the availability constraints that are associated with an expert sample.

\begin{table*}[h]
\caption{Demographics, role, and company details of the interview sample. }
\label{tab:interviewees}
\resizebox{.8\linewidth}{!}{%
\begin{tabular}{rllllll}
\multicolumn{1}{l}{} &
  \textbf{Role} &
  \textbf{Industry} &
  \textbf{Company Size} &
  \textbf{Gender} &
  \textbf{Age} &
  \textbf{Country of Residence} \\ \hline
I1 &
  \begin{tabular}[c]{@{}l@{}}Board / Management, \\ Sales, Customer Support\end{tabular} &
  \begin{tabular}[c]{@{}l@{}}Sales; Technology and \\ software development; \\ Pharmaceuticals\end{tabular} &
  \textgreater 25 000 &
  Male &
  55-64 &
  Switzerland \\ \hline
I2 &
  Board / Management &
  Public sector &
  \textless 25 &
  Male &
  25-34 &
  Germany \\ \hline
I3 &
  \begin{tabular}[c]{@{}l@{}}Technical, Customer \\ Support\end{tabular} &
  \begin{tabular}[c]{@{}l@{}}Admin or support; \\ Information; Logistics\end{tabular} &
  5 000 - 10 000 &
  Male &
  45-54 &
  United Kingdom \\ \hline
I4 &
  Technical &
  \begin{tabular}[c]{@{}l@{}}Technology and software \\ development\end{tabular} &
  10 000 - 25 000 &
  Male &
  35-44 &
  United Kingdom \\ \hline
I5 &
  Design &
  \begin{tabular}[c]{@{}l@{}}Technology and software \\ development; Healthcare and \\ Social Assistance; Education\end{tabular} &
  25 - 100 &
  Male &
  25-34 &
  South Africa \\ \hline
I6 &
  UX Researcher &
  \begin{tabular}[c]{@{}l@{}}Scientific services and research; \\ Technology and software \\ development; Healthcare and \\ Social Assistance\end{tabular} &
  \textless 25 &
  Female &
  25-34 &
  United States \\ \hline
I7 &
  Product &
  \begin{tabular}[c]{@{}l@{}}Sales; Information; Professional \\ services: Consulting or Advising\end{tabular} &
  25 - 100 &
  Male &
  18-24 &
  United Kingdom \\ \hline
I8 &
  \begin{tabular}[c]{@{}l@{}}Product, Researcher, \\ (UX) Researcher, Design\end{tabular} &
  Transportation &
  \begin{tabular}[c]{@{}l@{}}1 000 - \\ 5 000\end{tabular} &
  Male &
  25-34 &
  United Kingdom \\ \hline
I9 &
  Technical, Researcher &
  Manufacturing &
  100 - 500 &
  Male &
  25-34 &
  United Kingdom \\ \hline
I10 &
  CTO &
  \begin{tabular}[c]{@{}l@{}}Technology and software \\ development\end{tabular} &
  \textless 25 &
  Male &
  18-24 &
  United Kingdom \\ \hline
\end{tabular}%
}
\end{table*}

\section{Appendix C - Survey Results}

\subsection{Descriptive Results}
\label{appendix:surveyresults}
The following graphs provide further insights into the survey results discussed in \S\ref{sec:practitioner}.

\begin{figure*}[h]
  \includegraphics[width=.7\linewidth]{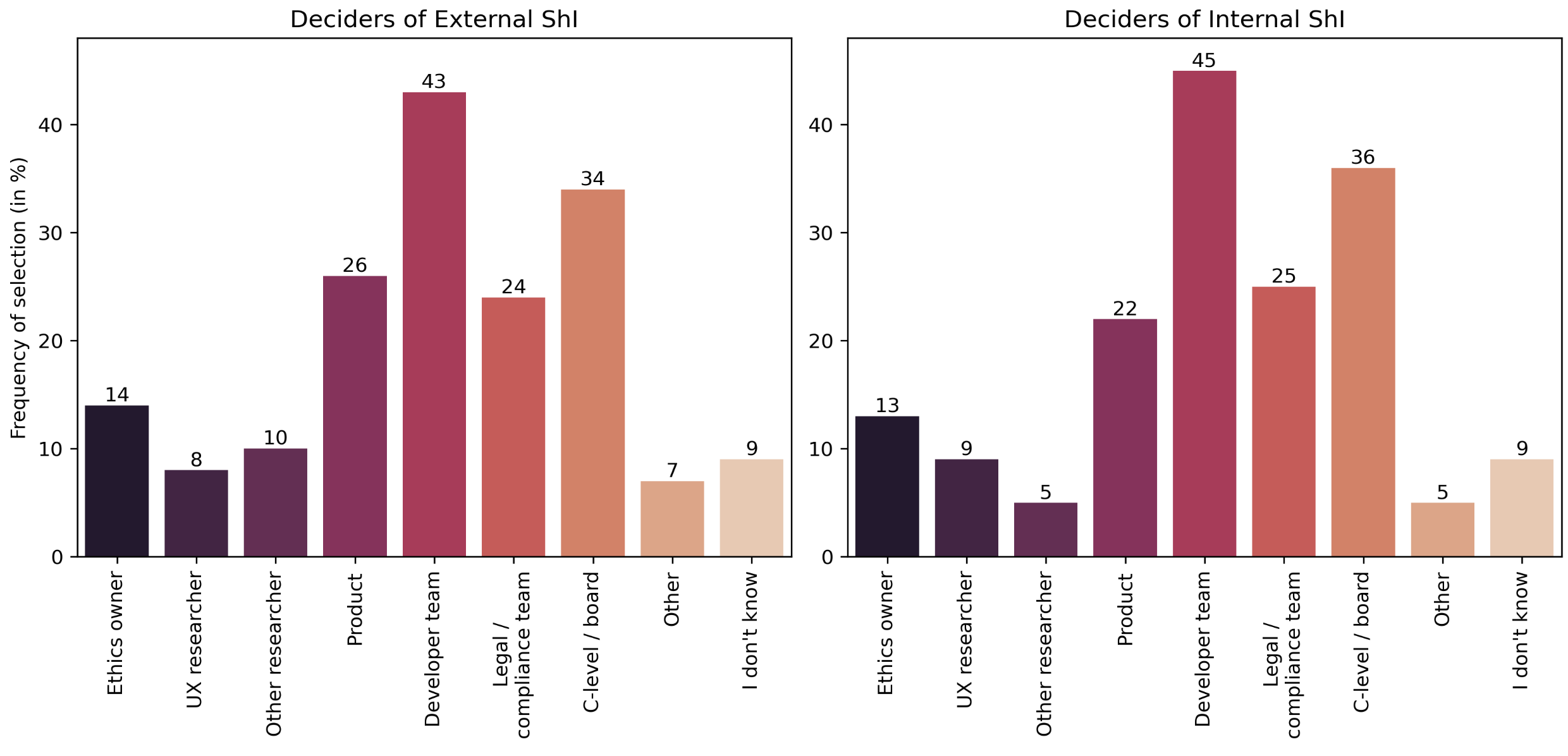}
  \caption{Answers of survey participants when asked who is deciding on external and internal SHI. The finding that the developer team, the C-suite, as well as the legal team are the main deciders, both for internal as well as external SHI, reflects the commercial and compliance-related drivers behind current SHI practices. }
  \label{fig:deciders}
  \Description{Answers of survey participants when asked who is deciding on external and internal SHI. Interestingly, developers and the management level are the main deciders for both types of stakeholders.}
\end{figure*}

\begin{figure*}[h]
  \includegraphics[width=.5\linewidth]{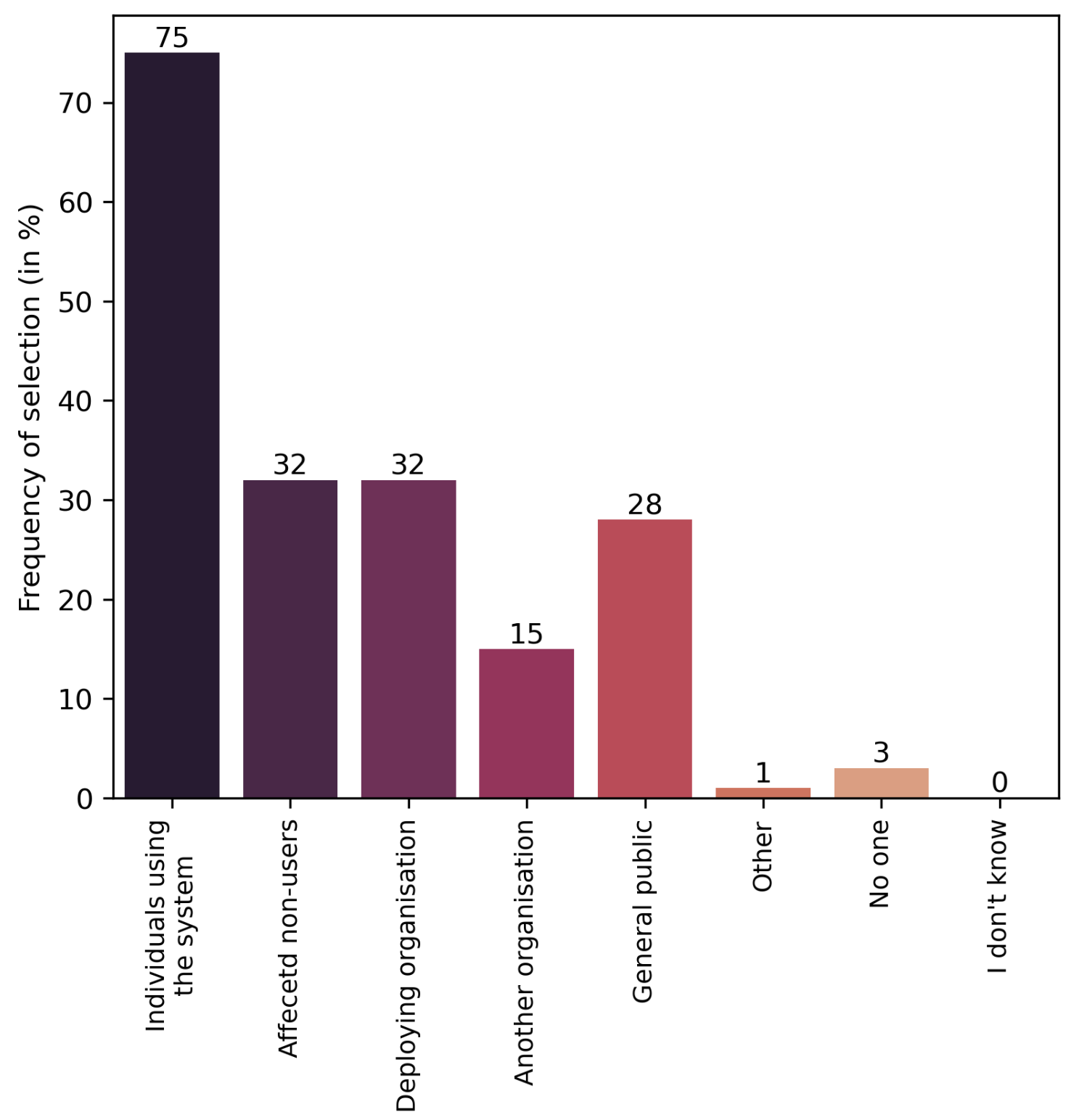}
  \caption{Answers of survey participants when asked who they expect to be impacted by the system they have been\slash are developing. For the groups \textit{affected non-users} and \textit{the general public}, around 30\% of participants expected impact. This pattern is in conflict with who was involved during development (4\% the general public,  11\% affected non-users).}
  \label{fig:impacted}
  \Description{Answers of survey participants when asked who they expect to be impacted by the system they have been\slash are developing. Around a third expects that non-users and the general public will be affected which is in stark contrast to who is involved.}
\end{figure*}

\begin{figure*}[h]
  \includegraphics[width=.8\linewidth]{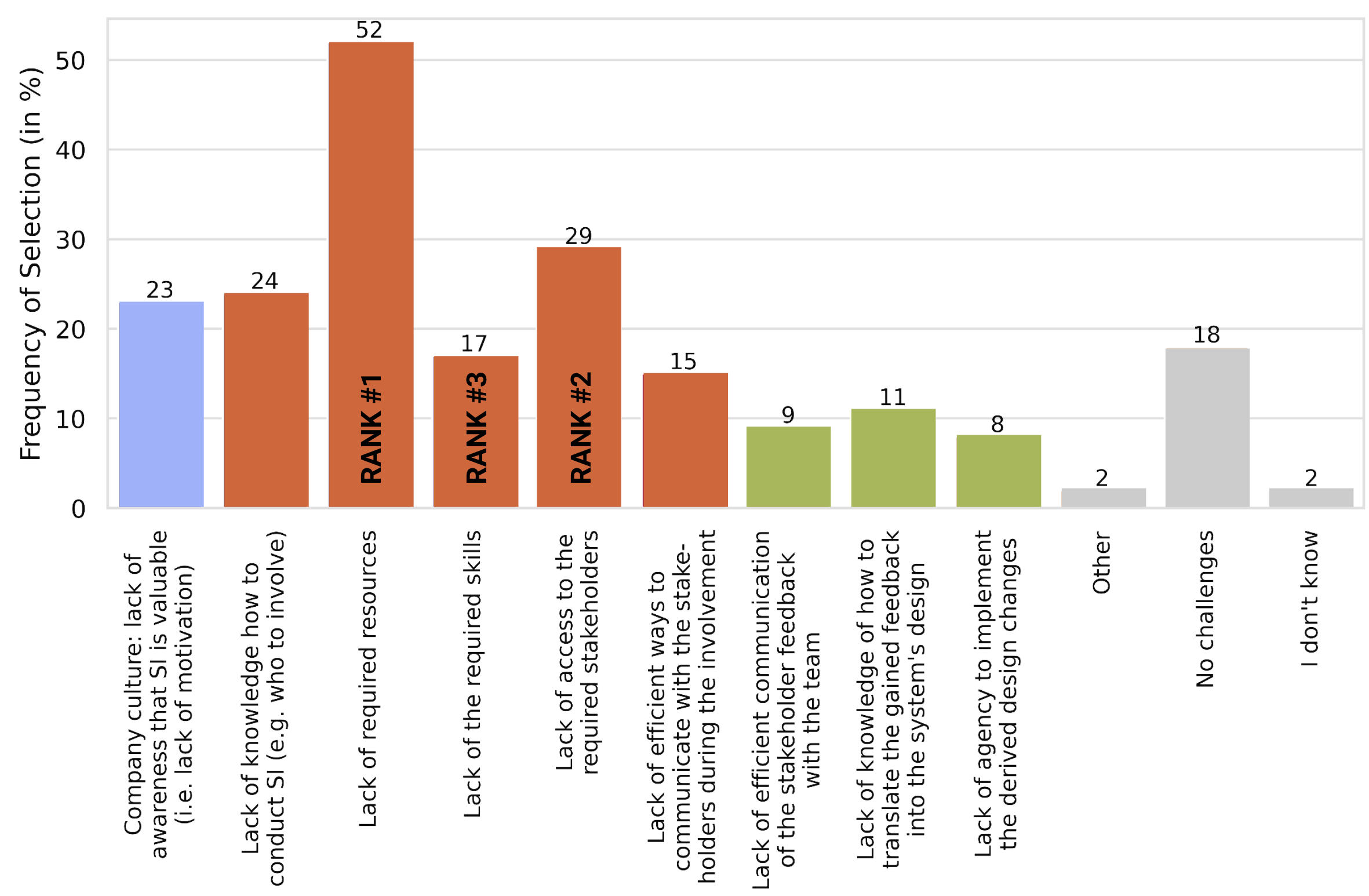}
  \caption{Challenges selected by survey participants, as well as the top three highest ranked challenges. In average, survey participants selected 1.73 challenges.}
  \Description{Bar chart illustrating the challenges selected by the survey participants. Lack of resources and access were the most selected and highest ranked challenges.}
\end{figure*}

\clearpage 

\subsection{Statistical Analysis}
\label{appendix:stats}

The following section provides an overview over the statistical tests ran against the survey results. Table \ref{tab:statsimpacttests} shows whether stakeholders that are expected to be impacted were more likely to be involved and Table \ref{tab:regulation} compares the scope of SHI between systems developed for regulated vs non-regulated industries.
Independent sample T-tests were Bonferroni corrected to account for multiple comparisons following \citet{weisstein2004bonferroni}. 

\begin{table*}[h]
\caption{T-Tests to test whether a specific stakeholder group was more likely to be involved if they were expected to be impact by the system than when no impact on the group was reported}
\label{tab:statsimpacttests}
\resizebox{\linewidth}{!}{%
\begin{tabular}{llll}
\hline
\textbf{impacted external sh} & \textbf{involvement when impacted} & \textbf{involvement when NOT impacted} & \textbf{comparison: T-tests} \\ \hline
\textbf{end-users} &
  \begin{tabular}[c]{@{}l@{}}\textit{typical end-users}\\ \textit{M} = 61.86\% (\textit{SD} = 48.83\%) \\ \textit{at-risk end-users}\\ \textit{M} = 8.25\% (\textit{SD} = 27.65\%)\end{tabular} &
  \begin{tabular}[c]{@{}l@{}}\textit{typical end-users}\\ \textit{M} = 45.45\% (\textit{SD} = 50.56)\\ \textit{at-risk end-users}\\ \textit{M} = 9.09\% (\textit{SD} = 29.19)\end{tabular} &
  \begin{tabular}[c]{@{}l@{}}\textit{typical end-users}\\ \textit{t}(128) = 1.7, \textit{p} = .101, \textbf{n.s.}\\ \textit{at-risk end-users}\\ \textit{t}(128) = -0.1, \textit{p} = .882, \textbf{n.s.}\end{tabular} \\ \hline
\textbf{impacted non-users}            & \textit{M} = 14.29\% (\textit{SD} = 35.42\%)         & \textit{M} = 9.09\% (\textit{SD} = 28.91).               & \textit{t}(128) = 0.9, \textit{p} = .375, \textbf{n.s. } \\ \hline
\textbf{the general public}            & \textit{M} = 8.11\% (\textit{SD} = 27.67\%)          & \textit{M} = 2.15\% (\textit{SD} = 14.58)                & \textit{t}(128) = 1.6, \textit{p} = .113, \textbf{n.s.} \\ \hline
\end{tabular}%
}
\end{table*}

\begin{table*}[h]
\caption{Independent sample T-tests to test whether the involvement of stakeholder groups differed between systems developed for regulated industries and systems developed for unregulated industries. Bonferroni corrected for multiple comparisons \citep{weisstein2004bonferroni}, i.e. \textit{p} threshold is .05 / 15 = .003. *\textit{significant after Bonferroni correction}, \textit{p} < .003; ***\textit{highly significant}, i.e. \textit{p} < .001 (°\textit{significant before Bonferroni correction (\textit{p} < .05) but not after}).}
\label{tab:regulation}
\resizebox{\linewidth}{!}{%
\begin{tabular}{llll}
\hline
\textbf{\begin{tabular}[c]{@{}l@{}}involved \\ stakeholders\end{tabular}} &
  \textbf{\begin{tabular}[c]{@{}l@{}}involvement of stakeholder \\ in regulated industry\end{tabular}} &
  \textbf{\begin{tabular}[c]{@{}l@{}}involvement of stakeholder \\ in NOT regulated industry\end{tabular}} &
  \textbf{\begin{tabular}[c]{@{}l@{}}t-test comparing \\ column B and C\end{tabular}} \\ \hline
\textbf{\begin{tabular}[c]{@{}l@{}}legal / compliance \\ team\end{tabular}} &
  \textit{M} = 58.62\% (\textit{SD} = 49.68\%) &
  \textit{M} = 29.03\% (\textit{SD} = 45.76) &
  \textit{t}(118) = 3.4, \textit{p} = \textbf{.001***} \\ \hline
\textbf{marketing} &
  \textit{M} = 36.21\% (SD = 48.48\%) &
  \textit{M} = 30.65\% (SD = 46.48) &
  \textit{t}(118) = 0.6, \textit{p} = .522, \textbf{n.s.} \\ \hline
\textbf{C-level / board} &
  \textit{M} = 39.66\% (\textit{SD} = 49.35\%) &
  \textit{M} = 20.97\% (\textit{SD} = 41.04) &
  \textit{t}(118) = 2.3, \textit{p} = .026°, \textbf{n.s.} \\ \hline
\textbf{product} &
  \textit{M} = 32.76\% (\textit{SD} = 47.34\%) &
  \textit{M} = 30.65\% (\textit{SD} = 46.48) &
  \textit{t}(118) = 0.2, \textit{p} = .806, \textbf{n.s.} \\ \hline
\textbf{HR} &
  \textit{M} = 12.07\% (\textit{SD} = 32.86\%) &
  \textit{M} = 8.06\% (\textit{SD} = 27.45) &
  \textit{t}(118) = 0.7, \textit{p} = .469, \textbf{n.s.} \\ \hline
\textbf{\begin{tabular}[c]{@{}l@{}}teams developing \\ other AI system\end{tabular}} &
  \textit{M} = 32.76\% (\textit{SD} = 47.34\%) &
  \textit{M} = 27.42\% (\textit{SD} = 44.97) &
  \textit{t}(118) = 0.6, \textit{p} = .528, \textbf{n.s.} \\ \hline
\textbf{no internal sh} &
  \textit{M} = 13.79\% (\textit{SD} = 34.78\%) &
  \textit{M} = 24.19\% (\textit{SD} = 43.18) &
  \textit{t}(118)=-1.4, \textit{p} = .151, \textbf{n.s.} \\ \hline
\textbf{typical end-users} &
  \textit{M} = 66.52\% (\textit{SD} = 47.95\%) &
  \textit{M} = 55.84\% (\textit{SD} = 50.17) &
  \textit{t}(118)=1.2, \textit{p} = .236,\textbf{ n.s.} \\ \hline
\textbf{at-risk end-users} &
  \textit{M} = 12.07\% (\textit{SD} = 32.86\%) &
  \textit{M} = 6.45\% (\textit{SD} = 24.77) &
  \textit{t}(118)=1.1, \textit{p} = .291, \textbf{n.s.} \\ \hline
\textbf{impacted non-users} &
  \textit{M} = 17.24\% (\textit{SD} = 38.10\%) &
  \textit{M} = 4.84\% (\textit{SD} = 21.63) &
  \textit{t}(118)=2.2, \textit{p} = .029°, \textbf{n.s.} \\ \hline
\textbf{privacy experts} &
  \textit{M} = 22.41\% (\textit{SD} = 42.07\%) &
  \textit{M} = 4.84\% (\textit{SD} = 21.63) &
  \textit{t}(118)=2.9, \textit{p} = .004°, \textbf{n.s.} \\ \hline
\textbf{domain experts} &
  \textit{M} = 31.03\% (\textit{SD} = 46.67\%) &
  \textit{M} = 16.13\% (\textit{SD} = 37.08) &
  \textit{t}(118)=1.9, \textit{p} = .054, \textbf{n.s.} \\ \hline
\textbf{\begin{tabular}[c]{@{}l@{}}external risk \\ assessment party\end{tabular}} &
  \textit{M} = 17.24\% (\textit{SD} = 38.10\%) &
  \textit{M} = 1.61\% (\textit{SD} = 12.70) &
  \textit{t}(118)=3.1, \textit{p} = \textbf{.003*} \\ \hline
\textbf{the general public} &
  \textit{M} = 3.44\% (\textit{SD} = 18.41\%) &
  \textit{M} = 4.84\% (\textit{SD} = 21.63) &
  \textit{t}(118)=-0.4, \textit{p} = .706, \textbf{n.s.} \\ \hline
\textbf{no external sh} &
  \textit{M} = 18.97\% (\textit{SD} = 39.55\%) &
  \textit{M} = 33.87\% (\textit{SD} = 47.71) &
  \textit{t}(118)=-1.9, \textit{p} = .066, \textbf{n.s.} \\ \hline
\end{tabular}%
}
\end{table*}

\end{document}